\documentclass[10pt,journal,compsoc]{IEEEtran}
\usepackage{algorithmic}
\usepackage{algorithm}
\usepackage{multirow}
\usepackage{multicol}
\usepackage{amssymb}
\usepackage{pifont}
\usepackage{booktabs}
\usepackage{bm}
\usepackage{makecell}
\usepackage{caption}
\usepackage{float} 
\usepackage{subfigure}
\usepackage{threeparttable}

\ifCLASSINFOpdf
   \usepackage[pdftex]{graphicx}
   \graphicspath{{../pdf/}{../jpeg/}}
   \DeclareGraphicsExtensions{.pdf,.jpeg,.png}
\else
   \usepackage[dvips]{graphicx}
   \graphicspath{{../eps/}}
   \DeclareGraphicsExtensions{.eps}
\fi
\usepackage{algorithmic}
\usepackage{algorithm}
\usepackage{multirow}
\usepackage{amssymb}
\usepackage{pifont}
\usepackage{booktabs}
\usepackage{bm}
\usepackage{array}
\begin{document}
%
\title{HSVRS: A Virtual Reality System of the Hide-and-Seek Game to Enhance Gaze Fixation Ability for Autistic Children}
%
%
%

\author{Chengyan~Yu, 
 Shihuan~Wang, 
 Dong~Zhang, 
 Yingying~Zhang,
 Chaoqun~Cen,
 Zhixiang~You,\\
 Xiaobing~Zou, 
 Hongzhu~Deng, 
 Ming~Li
\thanks{Chengyan Yu and Shihuan Wang contributed equally to this work.}
\thanks{This research is funded in part by the Science and Technology Program of Guangzhou City (202007030011), National Natural Science Foundation of China (62171207, 81873801, 62173353) and DKU Synear and Wang-Cai Seed Grant. (Corresponding author: Hongzhu Deng, Ming Li)
}
\thanks{
This work involved human subjects in its research. This study was approved by the Third Affiliated Hospital of Sun Yat-sen University Institutional Review Board (IRB No. [2020]02-118-03) and Duke Kunshan University Institutional Review Board (IRB No. 2022ML065).
}
\thanks{Chengyan Yu and Dong Zhang are with the School of Electronics and Information Technology, Sun Yat-sen University, Guangzhou 510006, China (e-mail: yuchy33@mail2.sysu.edu.cn; zhangd@mail.sysu.edu.cn).
}
\thanks{Ming Li is with the Data Science Research Center, Duke Kunshan University, Kunshan, China, 215316, (e-mail: ming.li369@dukekunshan.edu.cn).
}
\thanks{
Shihuan Wang, Yingying Zhang, Chaoqun Cen, Xiaobing Zou, Hongzhu Deng are with the Child Development and Behavior Center, Third Affiliated Hospital of Sun Yat-sen University,
No.600 Tianhe Road, Guangzhou, China, 510630, 
(email: wangshh53@mail.sysu.edu.cn; summer09251228@163.com;  cenchq@mail.sysu.edu.cn; zouxb@vip.tom.com; denghzh@mail.sysu.edu.cn).
}
\thanks{
Zhixiang You is with JazzVision Tech Co, Science and Technology City, Nanhu District, Jiaxing, Zhejiang, China, 314016, (email: zhx.you@gmail.com) 
}
}

\maketitle

\begin{abstract}
Numerous children diagnosed with Autism Spectrum Disorder (ASD) exhibit abnormal eye gaze pattern in communication and social interaction. Due to the high cost of ASD interventions and a shortage of professional therapists, researchers have explored the use of virtual reality (VR) systems as a supplementary intervention for autistic children. This paper presents the design of a novel VR-based system called the Hide and Seek Virtual Reality System (HSVRS). The HSVRS allows children with ASD to enhance their ocular gaze abilities while engaging in a hide-and-seek game with a virtual avatar. By employing face and voice manipulation technology, the HSVRS provides the option to customize the appearance and voice of the avatar, making it resemble someone familiar to the child, such as their parents. We conducted a pilot study at the Third Affiliated Hospital of Sun Yat-sen University, China, to evaluate the feasibility of HSVRS as an auxiliary intervention for children with autism (N=24). 
Through the analysis of subjective questionnaires completed by the participants' parents and objective eye gaze data, we observed that children in the VR-assisted intervention group demonstrated better performance compared to those in the control group. Furthermore, our findings indicate that the utilization of face and voice manipulation techniques to personalize avatars in hide-and-seek games can enhance the efficiency and effectiveness of the system.
\end{abstract}


%
\IEEEpeerreviewmaketitle

\section{Introduction}
\IEEEPARstart{A}{utism} Spectrum Disorder (ASD) is a neurodevelopmental disorder characterized by impairment in communication and social interaction, restricted interests and stereotyped behaviors\cite{lord2018autism,american2013dsm}. The prevalence of ASD has substantially increased in recent decades\cite{wallace2012global}. Currently, the overall prevalence of ASD among eight-year-old children in the United States has been reported to be approximately 1 in 44 children (2.27\%)\cite{maenner2021prevalence}. Abnormal eye gaze patterns are early emerging symptoms of ASD\cite{webb2014motivation}, and they have been incorporated into many diagnostic criteria and instruments.
Previous studies have shown that individuals with ASD tend to exhibit reduced gaze attention towards the upper face region, which contains social and emotional information\cite{chevallier2015measuring}, and increased gaze attention towards non-social objects or items.
Previous eye-tracking studies have also shown that children with ASD are more prone to gaze fixation deficits compared to typically developing individuals and those with other developmental disabilities, thus highlighting abnormal gaze patterns as critical characteristics of ASD\cite{jones2013attention,frazier2017meta}.

As said by the convention wisdom that "the earlier, the better", starting early intervention is critical in treating individuals with ASD\cite{towle2020earlier,lord2022lancet}. 
Currently, the mainstream modes of treatment for ASD involve behavioral interventions conducted by professional therapists and trained parents, including Applied Behavior Analysis (ABA)\cite{cooper2007applied} and the Early Start Denver Model (ESDM)\cite{dawson2010randomized}. 
These approaches typically involve manual training and require substantial human resources and financial investment from families\cite{carlon2013review}.
Due to the limited number of professional therapists specializing in ASD, children with ASD sometimes miss crucial opportunities for improvement due to long waiting times for intervention, particularly in developing countries where resources are scarce\cite{kasari2013interventions}.
In this context, digital applications hold significant potential to enhance social interaction and facilitate daily tasks for individuals with ASD in a cost-effective manner\cite{smith2008roles}.

\begin{figure}[!t]
\centering
\includegraphics[width=3.45in]{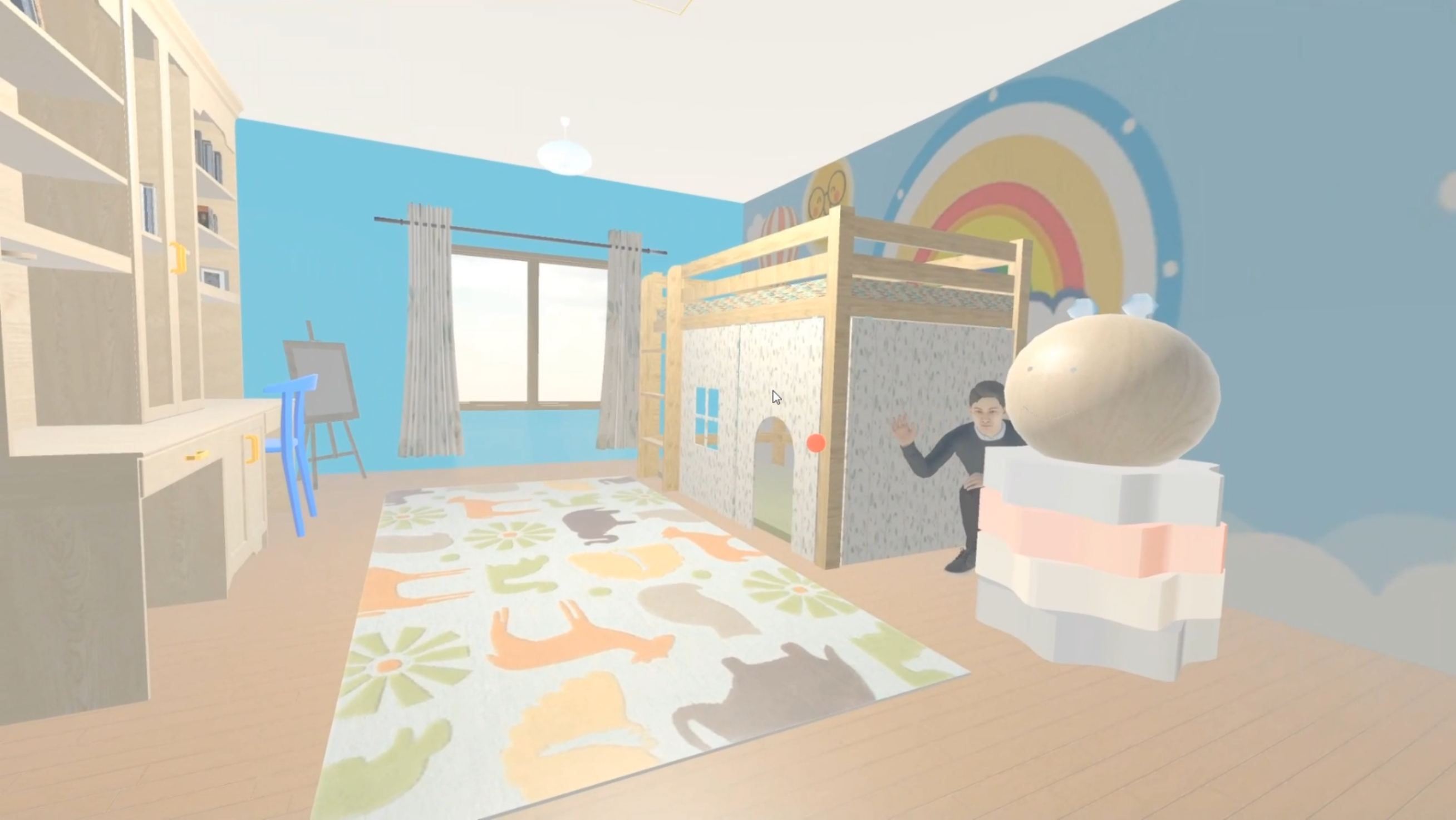}
\caption{The virtual environment in the HSVRS game is set in a room in which the parent avatar can hide in various locations. The parent avatar in this screenshot is hiding in front of the bed, waving at the user.}
\label{fig1}
\end{figure}

\begin{figure*}[!t]
\centering
\includegraphics[width=7in]{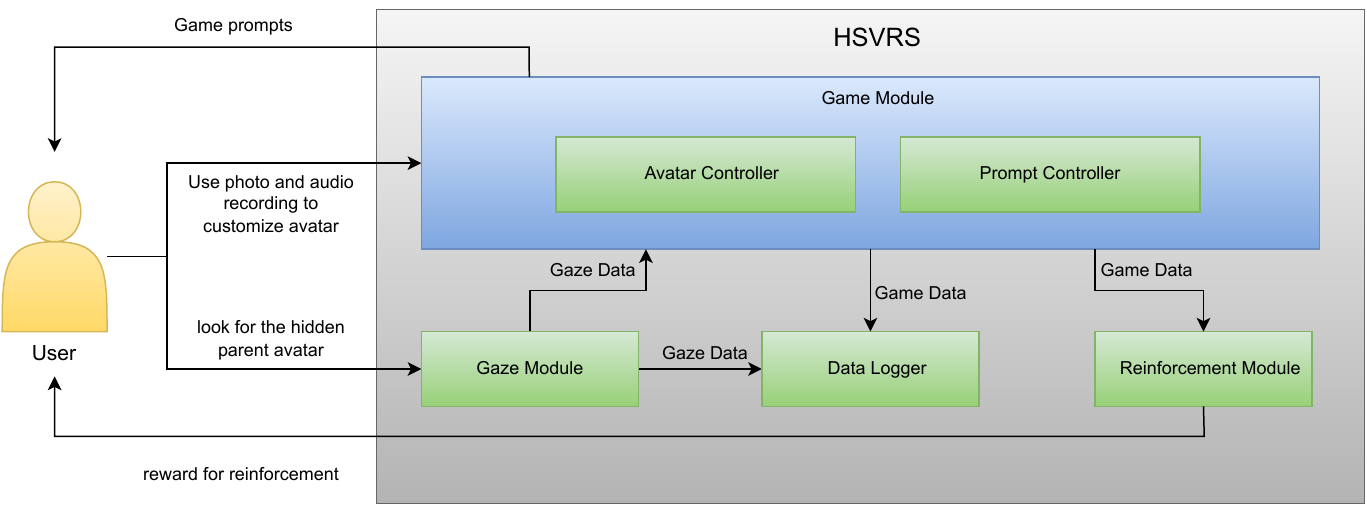}
\caption{Overview of the HSVRS system. Our HSVRS system consists of four components: Game Module (including Avatar Controller and Prompt Controller), Gaze Module, Data Logger and Reinforcement Module.}
\label{fig2}
\end{figure*}

Virtual reality (VR) technology with eye tracker has been increasingly applied in intervention for children with ASD\cite{bekteshi2020eye}. The eye tracker in the VR headset can serve as a control device, replacing the need for a regular joystick and thereby reducing interaction challenges. While virtual reality-based intervention cannot replace interventions conducted by professional therapists, it can provide a secure and immersive virtual environment in which children with ASD can engage with systems to enhance their social skills\cite{parsons2011state}. This study aims to investigate the use of VR technology to facilitate interventions for young children with ASD. Our hypothesis is that VR-based interventions can complement and augment regular interventions for young children with ASD.

In recent years, there has been a growing interest among researchers in utilizing virtual reality devices for assistive interventions in autism. Some researchers have focused on enhancing gaze fixation abilities\cite{berton2019studying,berton2020eye}, while others aim to improve joint attention skills\cite{yaneva2020detecting,courgeon2014joint,caruana2018joint,little2016gaze}, among other areas of investigation. The aforementioned studies demonstrate that advanced virtual reality technology has the potential to assist in the treatment of autism. However, only a limited number of studies have integrated VR with cutting-edge technologies, including deep learning-based computer vision and voice manipulation, to enhance conventional interventions. This study leverages deep learning-based computer vision and voice manipulation techniques to enable affordable customization of game characters within virtual reality devices. Specifically, we begin by pre-designing preliminary head and face shapes, and subsequently customize the avatar's face by modifying the UV texture using a deep learning-based model that is trained on specific photos. To customize the avatar's voice, we collect emotionally loaded video recordings.

In this study, we present a novel VR-based system called the Hide and Seek Virtual Reality System (HSVRS). The system is specifically designed to facilitate hide-and-seek games for children with Autism Spectrum Disorder (ASD) by utilizing customized virtual avatars, such as their parents. The primary aim of HSVRS is to improve the eye gaze ability of children with ASD through engaging gameplay.
HSVRS provides a virtual home scene equipped with furniture, creating a suitable environment for playing hide-and-seek. During gameplay, users simply need to turn their heads and utilize their eye gaze to locate the hidden avatar in order for the game to function properly. The system incorporates built-in eye-tracking devices, making it accessible and user-friendly for children of all age groups.
Significantly, the avatar within the system can be customized by uploading a facial photograph and a few audio recordings of the parent. This customization feature allows the avatar to closely resemble and sound like the child's actual parent, thereby potentially enhancing the transfer effects of the training.

\section{SYSTEM DESIGN}
We design the Hide and Seek Virtual Reality System (HSVRS) to help children practice their skills of gaze fixation. 
The central gameplay of HSVRS revolves around a hide-and-seek game, wherein the user interacts with an avatar representing their parents. Previous research has demonstrated the engaging nature of hide-and-seek games for children with Autism Spectrum Disorder (ASD) \cite{brodhead2014use}. Our system features a virtual family room scene with furniture, providing a realistic environment for the parent avatars to conceal themselves. Figure 1 presents a screenshot captured during actual gameplay, offering a glimpse into the immersive experience.
Playing hide-and-seek with VR headsets is straightforward as users can actively participate by controlling their eye gaze. A notable aspect of HSVRS is its capacity to customize the parent avatars using cutting-edge deep learning-based computer vision and voice conversion techniques. By providing a single photo and a few audio recordings, we can ensure that the avatar closely resembles the child's parent both in appearance and voice.

\subsection{Overview}

Fig. \ref{fig2} presents the interaction diagrams depicting the communication between the user and the components of HSVRS. The system comprises four key components: the Game Module (consisting of the Avatar Controller and Prompt Controller), Gaze Module, Data Logger, and Reinforcement Module.
Within the Game Module, the Avatar Controller utilizes a parent's photo and 80 audio utterances to customize the avatar. It manages the parent avatars, facilitating their hiding and transitioning between different hiding locations. Concurrently, the Prompt Controller receives gaze data from the Gaze Module, enabling it to provide prompts when the user struggles to locate the parent avatar. The collaborative efforts of the Avatar Controller and Prompt Controller ensure the smooth progression and functionality of the game.
The Gaze Module, equipped with a built-in eye tracker, captures the user's gaze data during the hide-and-seek gameplay. It relays this data to both the Game Module and the Data Logger for further processing and analysis. The Data Logger receives and stores both gaze data and game data, enabling subsequent analysis and evaluation.
The Reinforcement Module leverages the game data to determine whether the user has successfully found the parent avatar. In the event of a successful find, the Reinforcement Module provides rewards in the form of visual and auditory reinforcers. Visual reinforcers encompass captivating special effects, thumbs up gestures from the parents' avatars, and interactions with user-favorite cartoon characters. Auditory reinforcers include pleasant music and encouraging words spoken by the parent character, among other positive stimuli.

\subsection{Game Module}
The Game Module comprises two main components: the Avatar Controller and the Prompt Controller. This module assumes responsibility for managing the hiding behavior of the parent avatar within the virtual home scene. It facilitates the hiding process and dynamically alters the avatar's hiding location once it has been discovered. Additionally, in cases where the user struggles to locate the parent avatar, the Prompt Controller intervenes by offering helpful prompts to assist the user in the gameplay.

\begin{figure}[!t]
\centering
\includegraphics[width=3.4in]{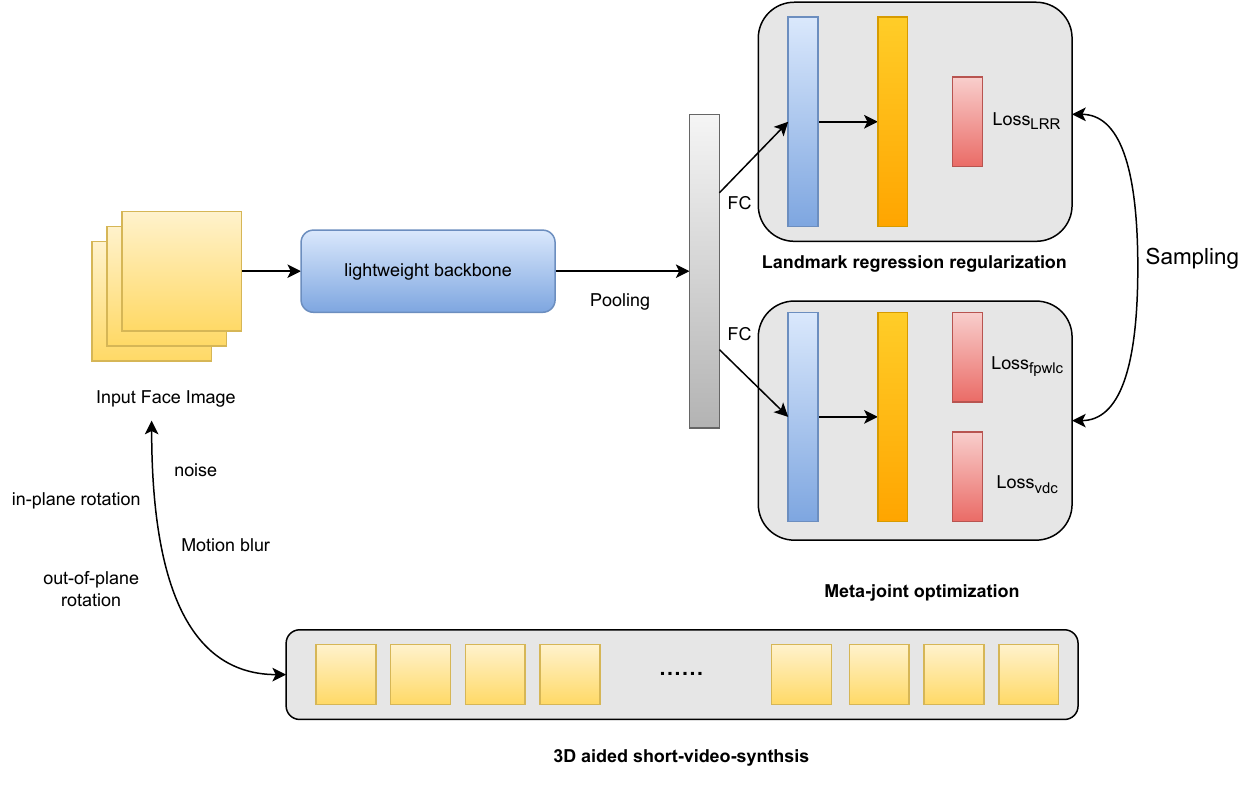}
\caption{The architecture of the 3DDFA\_V2 network.The 3DDFA\_V2 architecture consists of four parts: the lightweight backbone, the landmark-regression regularization, the meta-joint optimization of fWPDC and VDC\cite{zhu2016face,zhu2017face}, and the 3D-aided short-video-synthesis.}
\label{fig3}
\end{figure}
\begin{figure}[!t]
\centering
\includegraphics[width=3.4in]{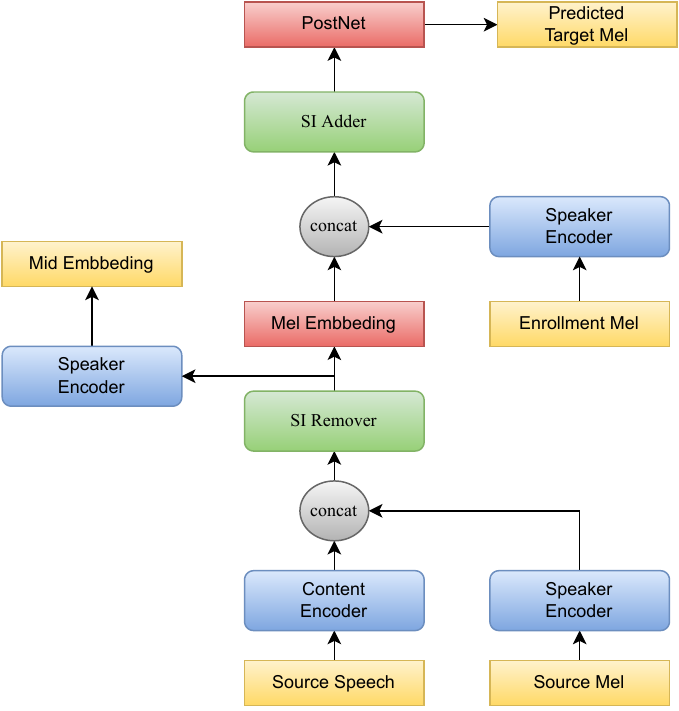}
\caption{The structure of the SIG-VC system, where SI stands for speaker information\cite{zhang2022sig}. SIG-VC proposes a supervised intermediate representation to reduce speaker information and obtain purer content representations.}
\label{SIGVC}
\end{figure}

\begin{figure}[!t]
\centering
\includegraphics[width=3.4in]{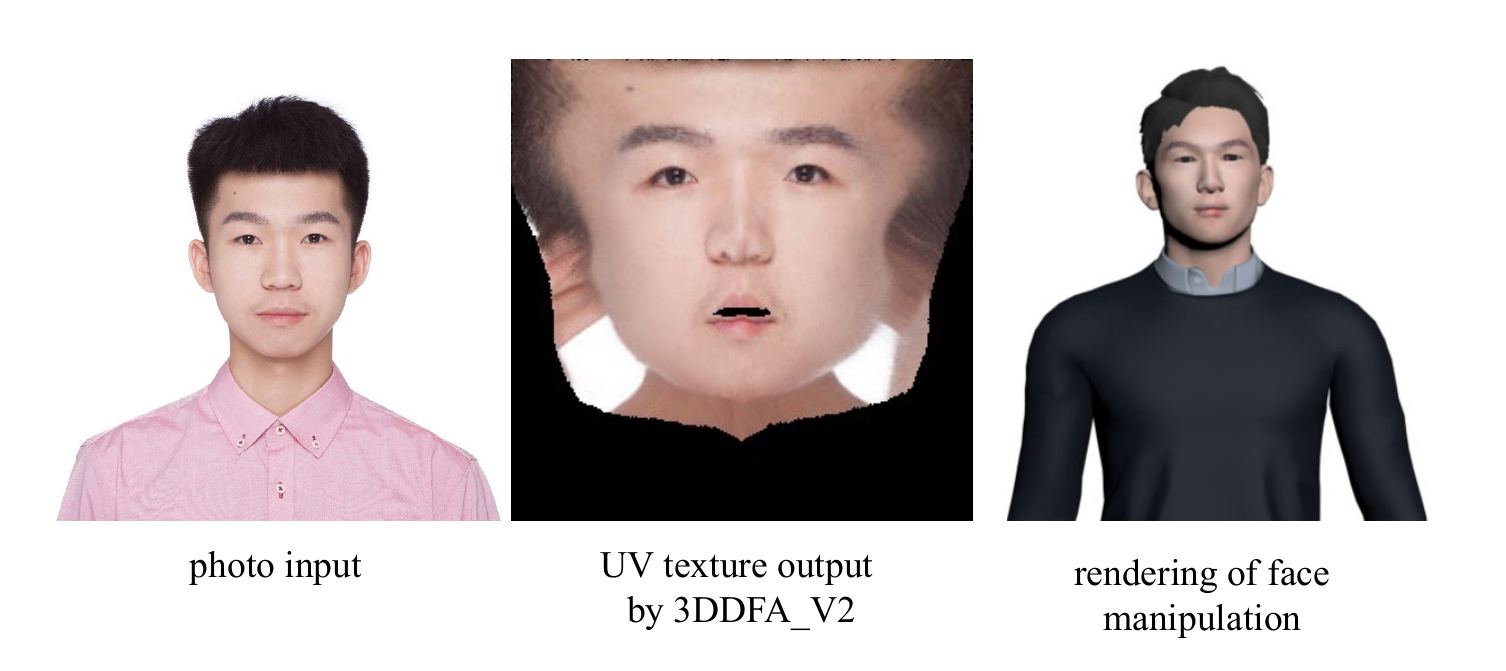}
\caption{An example of face manipulation.}
\label{facemanipulation}
\end{figure}

\subsubsection{Avatar Controller}

We employed the application 3D Studio Max for designing the animation of the parent avatars. In order to examine the potential impact of face and voice deepfake manipulation on the intervention effect, we created two distinct types of avatars: simulated avatars and customized avatars.
The simulated avatars feature a consistent face and voice across all instances. On the other hand, for the customized avatars, we utilized the photos and recordings provided by the participants to manipulate both the appearance and voices of the avatars. This allowed us to create personalized avatars that closely resemble and sound like the participants' actual parents.

We now describe the implementation details in the face manipulation module. We pre-designed some of the rough facial shapes, such as the large eyes, high nose, and tiny mouth. Moreover, we crafted a range of common hairstyles such as the rounded inch, the backward style, and the bald head. In 3D Studio Max, the character's face is represented by a UV texture, which forms the basis for our face manipulation process \cite{mullen2011mastering}.
For face manipulation, we employed an open-source implementation of 3DDFA\_V2 \cite{guo2020towards}. 3DDFA\_V2 is a deep learning-based algorithm that enables 3D dense face alignment. Its regression network framework strikes a balance between speed, accuracy, and stability. Fig. \ref{fig3} illustrates the architecture of 3DDFA\_V2, which comprises four main components: the lightweight backbone, landmark-regression regularization, meta-joint optimization of fWPDC and VDC \cite{zhu2016face, zhu2017face}, and 3D-aided short-video synthesis.

In the 3DDFA\_V2 architecture, input images are processed by the backbone network to extract features. These features are then passed through a pooling operation, generating two sets of tensors. One set of tensors is utilized to compute the landmark regression loss, while the other set is employed to calculate the Vertex Distance Cost (VDC) and Weighted Parameter Distance Cost (WPDC) \cite{zhu2016face, zhu2017face}. The meta-joint optimization effectively combines the advantages of both cost functions.
Additionally, the 3DDFA\_V2 algorithm introduces a 3D-aided short-video synthesis method. By leveraging this technique, the network simulates both in-plane and out-of-plane facial movements, thereby transforming a still image into a short video. This approach enhances the stability of videos constructed from limited training data comprising only still images.


\begin{figure}[!t]
\centering
\includegraphics[width=3in]{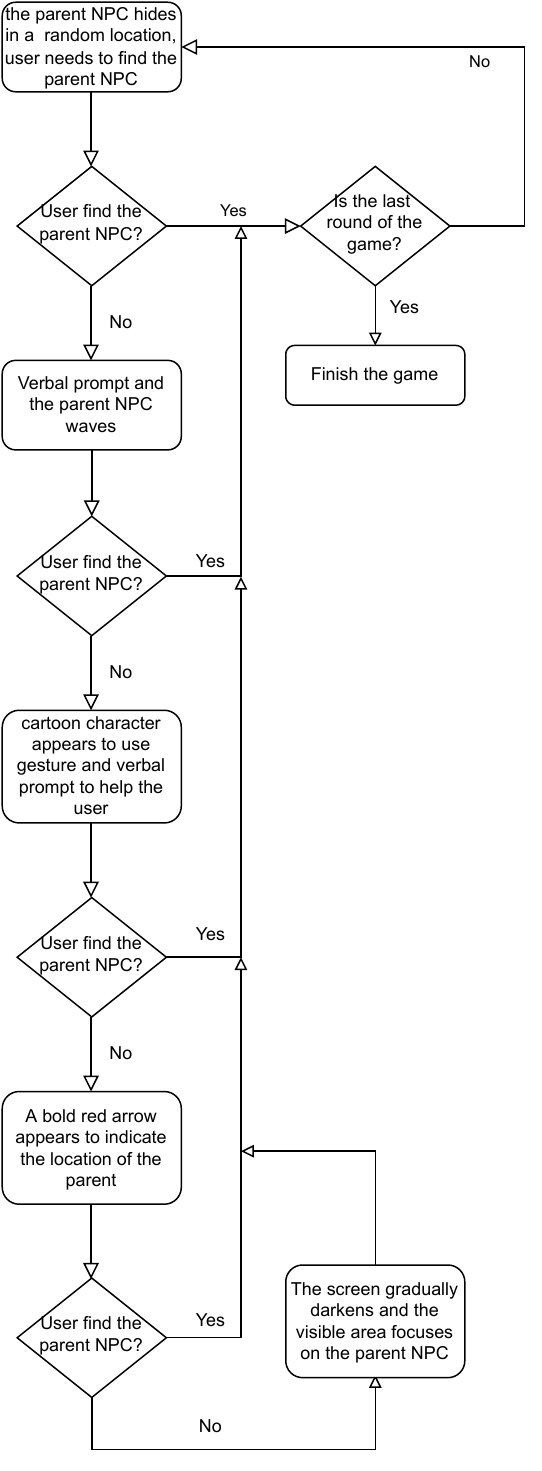}
\caption{Assistive flow chart in our prompt Controller. In the HSVRS system, we provide verbal, gestural and physical prompts. The prompts work together to guide participants to find the hidden parent avatar.}
\label{fig5}
\end{figure}

We now turn to the implementation details in the voice manipulation module. We invite a student speaker to record the voices of game dialog in an expressive manner, and collect 80 audio utterances from the user's parents. In terms of the voice conversion algorithm, we adopt the transformer\cite{vaswani2017attention} based Speaker Information Guided Voice Conversion system (SIG-VC)\cite{zhang2022sig} to convert the prerecorded expressive dialog voice into each parent's characteristics. The SIG-VC system employs a supervised intermediate representation to reduce source speaker information and obtain purer content representations. SIG-VC uses a pre-trained speaker verification system for speaker representation extraction, and applys a pre-trained acoustic model\cite{deliyski1993acoustic} to extract the linguistic feature. Moreover, this system enforces intermediate representations in the vector space of selected acoustic features by sharing parameters for certain modules. Finally, a feedback loss\cite{cai2020speaker,du2021optimizing} is applied to assign high spoofing capability to the generated speech. The SIG-VC demo\footnote{https://haydencaffrey.github.io/sigvc/index.html} is available online for readers.

\subsubsection{Prompt Controller}
The Prompt Controller plays a crucial role in assisting users who have difficulty finding the parent avatar. An effective instructional strategy for prompting is the least-to-most (LTM) prompting mechanism\cite{libby2008comparison}. 
The LTM mechanism has been widely utilized in previous research to teach various skills, including motor skills \cite{yanardaug2011effects} and communication skills \cite{finke2017effects,polick2012comparison}.
The LTM prompt mechanism employs a sequence of prompts to support the user during the game. When the Prompt Controller delivers an instruction, it utilizes a series of prompts, beginning with the least intrusive and progressively advancing to more instructive prompts. Prompts are presented individually until the user successfully finds the parent avatar. Typically, the LTM prompting mechanism employs a set of three consecutive prompts to facilitate the acquisition of a new skill. These prompts can include verbal cues, gestures, modeling, or physical assistance. In our system, the LTM mechanism allows the user to make their best effort to locate the hidden parent avatar before additional prompts are provided.

Table \ref{assistivePrompts} lists the assistive prompts in the hide-and-seek game. The level indicates the strength of the prompt. The larger the level, the more help the player gets. Fig. \ref{fig5} shows the flow chart of the assistive prompt.

\begin{table}[]
\centering
\caption{the assitive prompts in hide and seek game.}
\label{assistivePrompts}
\renewcommand{\arraystretch}{1.3}
\begin{tabular}{cc}
\hline
Intensity Level & Assitive Prompts                                                                                                              \\ \hline
1               & \begin{tabular}[c]{@{}c@{}}Voice prompts for avatar's location \end{tabular}                        \\ \hline
1               & \begin{tabular}[c]{@{}c@{}}The avatar will wave at the user\end{tabular}                                            \\ \hline
2 & \begin{tabular}[c]{@{}c@{}}A cartoon character appears to help \\ the user using voice and gesture prompts\end{tabular} \\ \hline
2               & \begin{tabular}[c]{@{}c@{}}A bold red arrow appears to  indicate \\the location of the avatar\end{tabular}         \\ \hline
3               & \begin{tabular}[c]{@{}c@{}}The screen gradually darkens, and \\the visible area focuses on the avatar\end{tabular} \\ \hline
\end{tabular}
\end{table}

\begin{figure}[h]
 \begin{minipage}{0.45\linewidth}
     \vspace{3pt}  
     \centerline{\includegraphics[width=\textwidth]{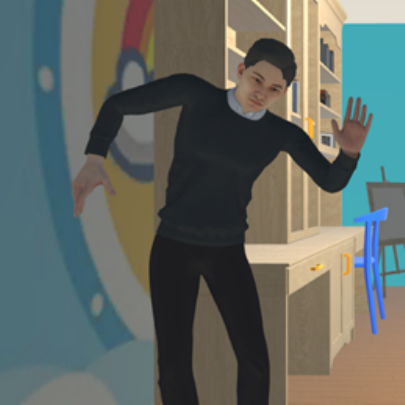}}
     \vspace{3pt}
     \centerline{\includegraphics[width=\textwidth]{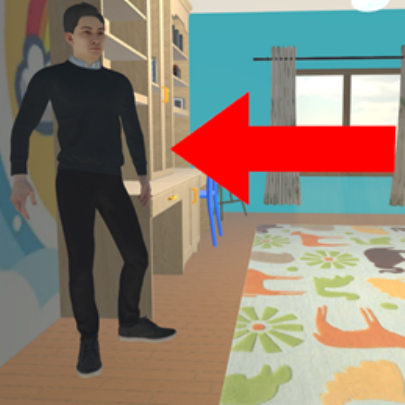}}
     \vspace{4pt}
 \end{minipage}
 \begin{minipage}{0.45\linewidth}
     \vspace{3pt}  
     \centerline{\includegraphics[width=\textwidth]{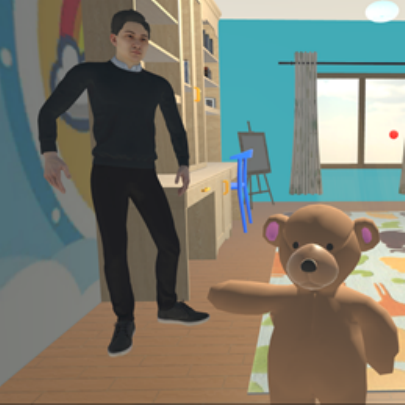}}
     \vspace{3pt}
     \centerline{\includegraphics[width=\textwidth]{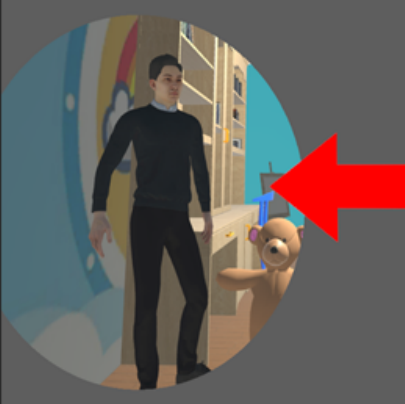}}
     \vspace{3pt}
 \end{minipage}
	\caption{visualization of the assitive prompts in the hide and seek game.}
	\label{fig7}
\end{figure}

\subsection{Gaze Module}
In this study, we have developed a Gaze Module that utilizes the built-in 7Invensun\footnote{https://haydencaffrey.github.io/sigvc/index.html} eye tracker to analyze gaze behavior in real-time. The eye tracker has a sampling frequency of 30 Hz and is integrated with the 7Invensun Unity development package. We collect continuous gaze coordinate data during the gameplay. Specifically, when the user's gaze remains within the avatar's body box in Unity for a duration of 500ms, the system determines that the player has successfully located the parent avatar and proceeds to the next round of the game.
Our main focus in this study is on fixation data. To facilitate data collection, we have predefined multiple regions of interest (ROIs) within Unity. Whenever the user's gaze point falls within these predefined ROIs, the Gaze Module records and sends the corresponding gaze data to the Data Logger. The ROIs are categorized into three types: face ROI, body ROI, and background ROI. The face ROI is defined by the avatar's facial region, which contains significant social and emotional information\cite{haxby2002human}. The body ROI encompasses the avatar's body, providing cues that participants are still attending to the parent avatar, albeit with fewer social cues than the face region. The background ROI represents various elements in the virtual scene, such as furniture, walls, and other objects, which offer minimal social cues.
During the gameplay, the Gaze Module captures the user's real-time gaze fixation data and transmits it to the Data Logger. Once the game is completed, the Gaze Module calculates the percentage of gaze fixations that occur within each ROI region throughout the game and forwards this information to the Data Logger for further analysis.

\begin{figure}[!t]
\centering
\includegraphics[width=3.4in]{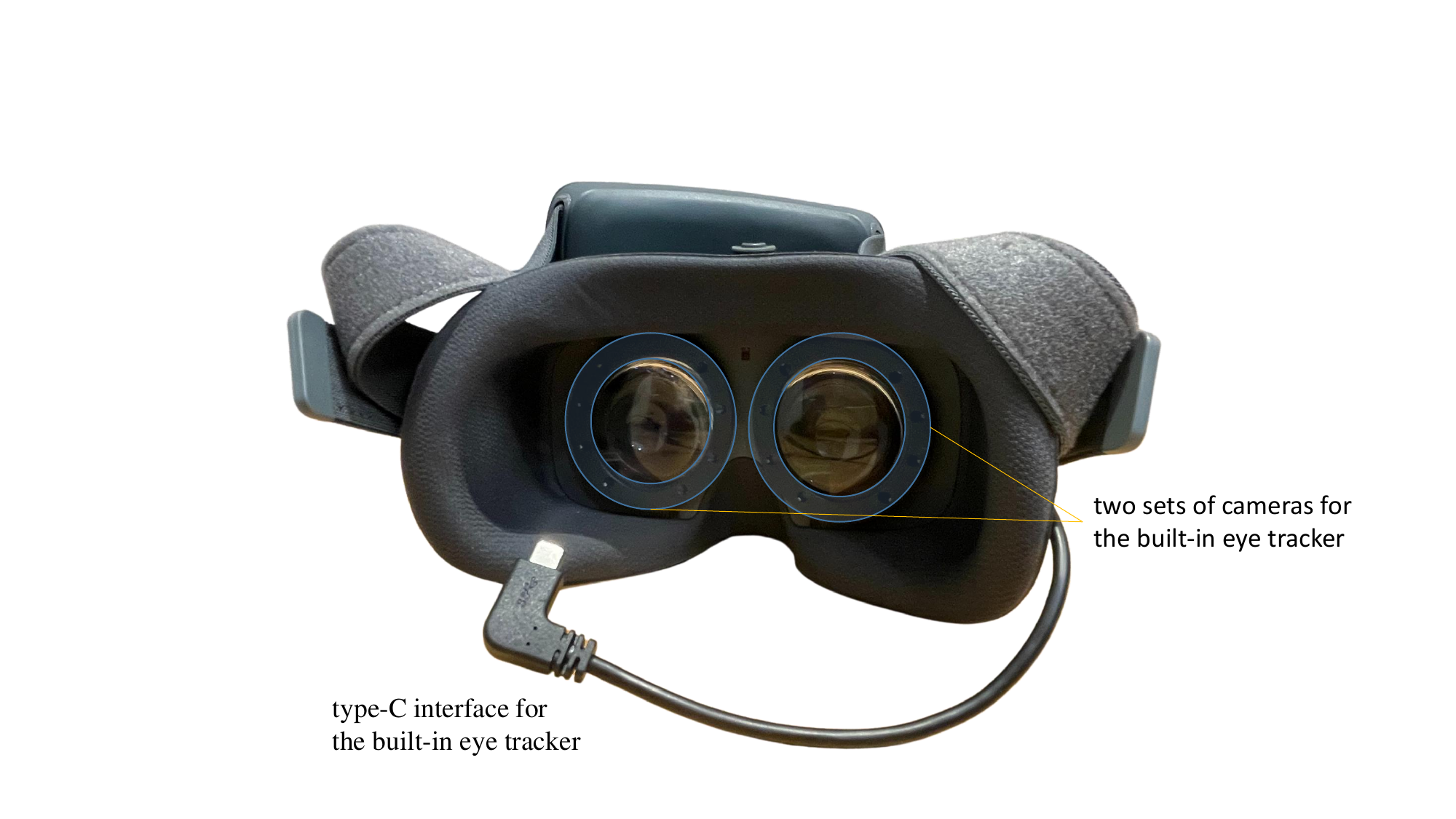}
\caption{The VR headset with its bulit-in gaze tracker.}
\label{fig8}
\end{figure}

\begin{figure}[!t]
\centering
\includegraphics[width=3in]{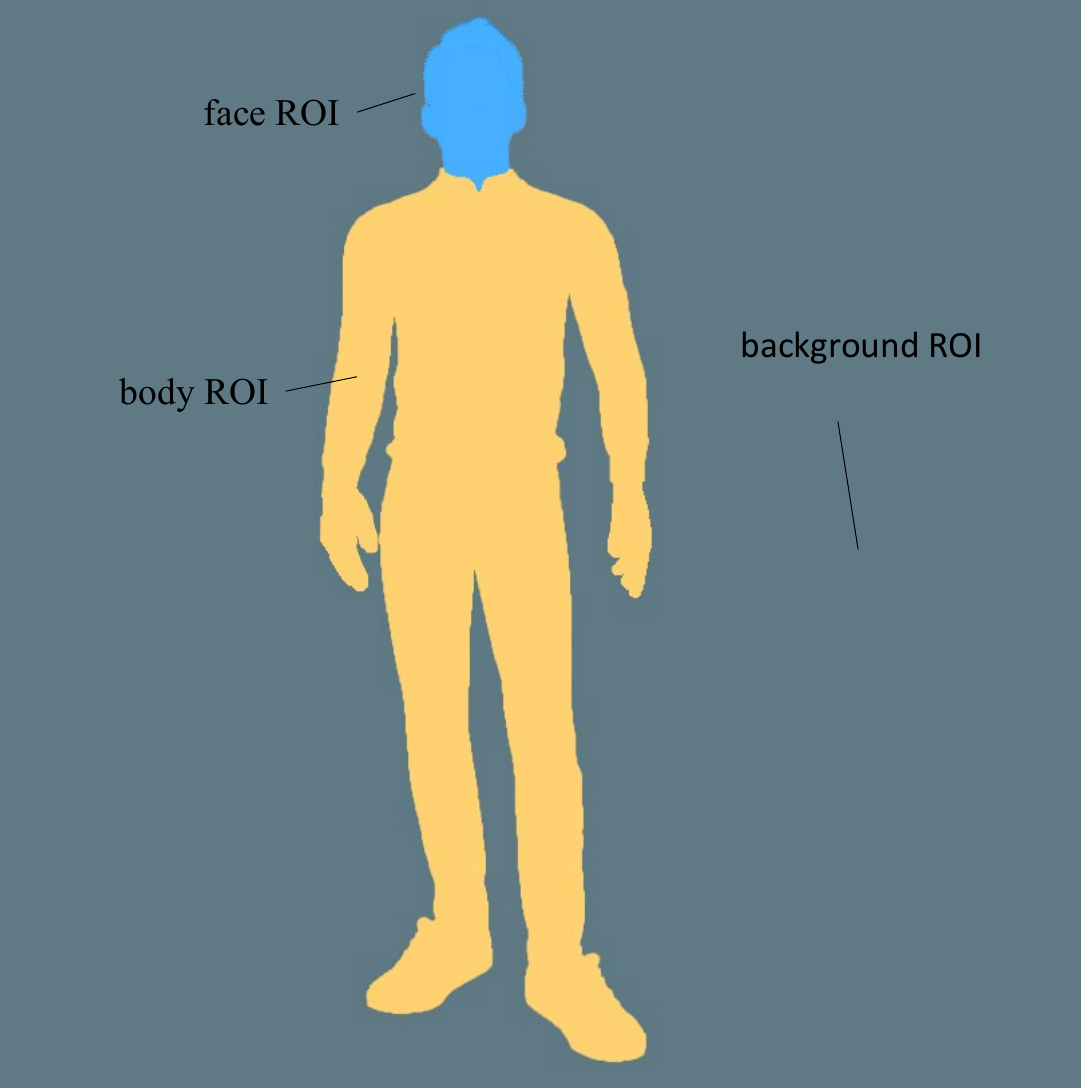}
\caption{The predefined three types of ROI: face ROI, body ROI and background ROI. Face ROI and body ROI are defined by avatar's face and body regions. The background ROI are defined by the scene's background.}
\label{fig9}
\end{figure}

\subsection{Reinforcement Module}
Reinforcement plays a crucial role in intervention systems, and it holds particular importance for children with autism\cite{williams1981response,kang2013effects}. An effective reinforcement system can greatly enhance the effectiveness of an intervention\cite{charlop1998using,koegel2009improving}. In the Hide and Seek Virtual Reality System (HSVRS), we have designed a comprehensive reinforcement system that offers various types of reinforcers to motivate and reward the children. Table \ref{reinforcers} outlines the reinforcers available in the HSVRS system.
The reinforcement system in HSVRS includes a wide range of visual and audio rewards. During the hide-and-seek game, when the child successfully finds the hiding location of the parent avatar, the Reinforcement Module provides verbal praise and thumbs-up gestures from the parent avatar as a form of positive reinforcement. Furthermore, in addition to the parent avatars, the virtual scenes are enriched with captivating special effects. To further engage the children, HSVRS features a collection of pre-designed cartoon characters from which the children can choose their favorite character before beginning the game. As the hide-and-seek game progresses and all the hiding locations are found, the Reinforcement Module displays the child's selected cartoon character. This character will offer words of encouragement and perform an extended dance routine, providing additional visual and audio reinforcement.

\subsection{Data Logger}
The data logger collects data in virtual environment, including real-time gaze data and time stamps of different events during the game.

\begin{table}[]
\caption{The reinforcers in our HSVRS system.}
\label{reinforcers}
\begin{tabular}{@{}cll@{}}
\toprule
\multicolumn{1}{l}{No.} &
  Reinforcer &
  \begin{tabular}[c]{@{}l@{}}When the reinforcement\\ arises\end{tabular} \\ \midrule
1 &
  \begin{tabular}[c]{@{}l@{}}verbal praise of \\ avatar characters\end{tabular} &
  \begin{tabular}[c]{@{}l@{}}The user finds \\ the parent avatar\end{tabular} \\ \midrule
2 &
  \begin{tabular}[c]{@{}l@{}}thumbs up from \\ the parent avatar\end{tabular} &
  \begin{tabular}[c]{@{}l@{}}The user finds \\ the parent avatar \end{tabular} \\ \midrule
3 &
  \begin{tabular}[c]{@{}l@{}}The parent avatar \\ performs a short dance\end{tabular} &
  \begin{tabular}[c]{@{}l@{}}The user finds \\ the parent avatar \end{tabular} \\ \midrule
4 &
  \begin{tabular}[c]{@{}l@{}}display random cool \\ effects like fireworks exploding\end{tabular} &
  \begin{tabular}[c]{@{}l@{}}The user finds \\ the parent avatar \end{tabular} \\ \midrule
\end{tabular}
\end{table}

\begin{figure}[!t]
\centering
\includegraphics[width=3.4in]{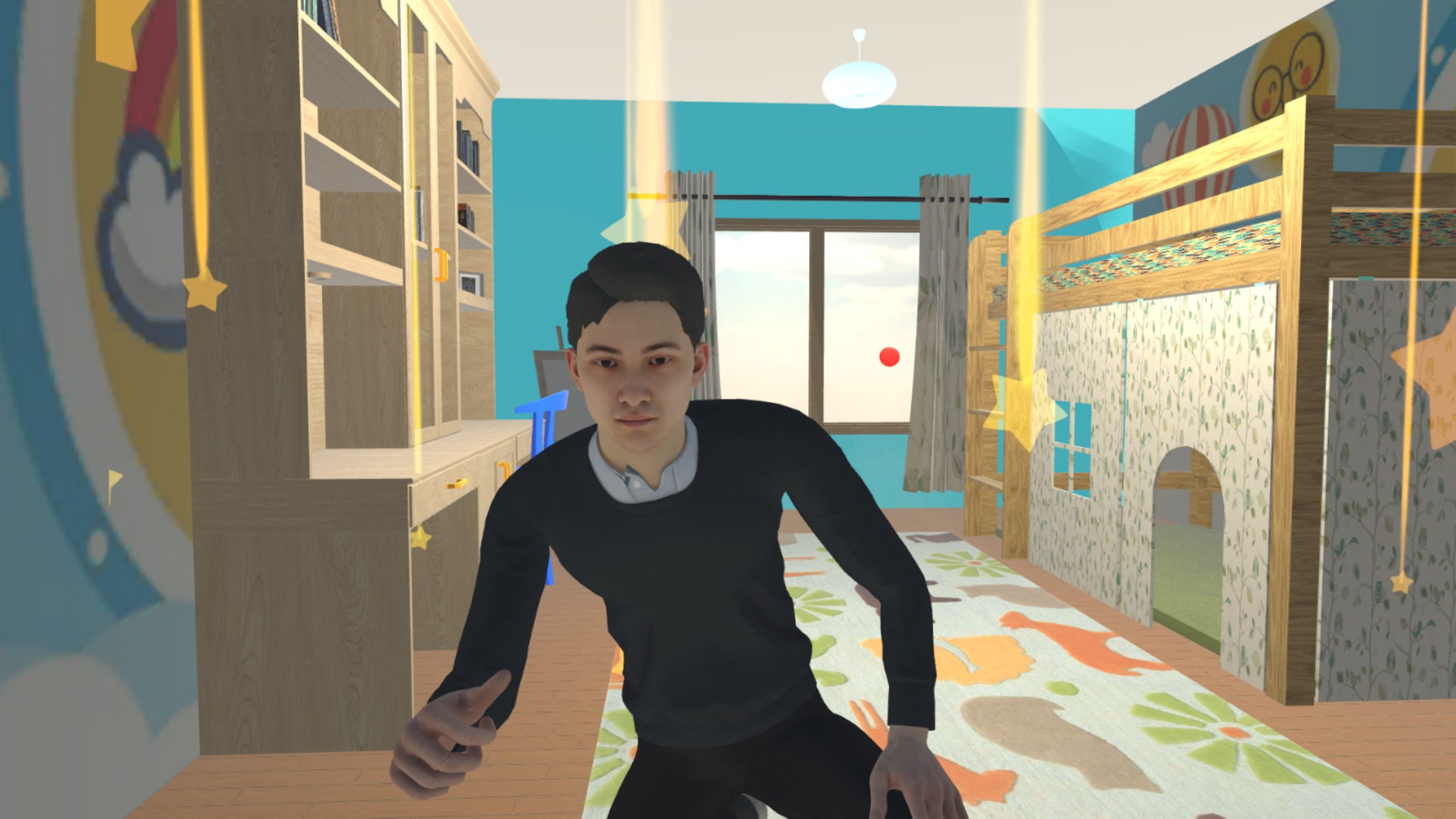}
\caption{An example of reinforcer in our HSVRS system.}
\label{fig10}
\end{figure}

\section{EXPERIMENTAL DESIGN}

We conducted a pilot study to evaluate the hypothesis that incorporating the Hide and Seek Virtual Reality System (HSVRS) as an ancillary intervention would lead to improvements in gaze fixation among children with autism. This hypothesis was assessed through subjective questionnaire analysis. Furthermore, we aimed to compare the intervention effects between two groups: the avatar customized group and the uncustomized group. We hypothesized that children from the avatar customized group would demonstrate superior performance compared to those from the uncustomized group, as evidenced by reduced completion time and average response time, as well as an increased game score. Additionally, we explored whether the inclusion of face and voice deepfake manipulation had an impact on the gaze patterns observed across different Regions of Interest (ROIs).

\subsection{Participants}

We recruited a total of twenty-four children with Autism Spectrum Disorder (ASD) from the Child Developmental \& Behavioural Center of the Third Affiliated Hospital of Sun Yat-sen University in Guangzhou. These children not only participated in our experiment but also received intervention training sessions at the hospital. They were divided into three groups as follows: 
\begin{itemize}
    \item Control Group: This group consists of children who solely participated in the intervention program at the hospital
    \item Avatar Uncustomized Group: In addition to the hospital intervention, children in this group utilized the HSVRS as an auxiliary intervention. The avatars used in the HSVRS were not customized for these children.
    \item Avatar Customized Group: Similar to the Avatar Uncustomized Group, children in this group received the hospital intervention and utilized the HSVRS as an auxiliary intervention. However, in this case, the avatars in the HSVRS were customized specifically for each child.
\end{itemize}

\subsection{Protocol}
In this study, we adopt a subjective questionnaire to the parents of all participants both before and after the experiment. Additionally, we conducted five repeated trials over a period of seven days for both the avatar uncustomized and avatar customized groups. During each visit, children from both groups were instructed to play a game of hide-and-seek using our HSVRS system. It is important to note that the hiding location for each round of the game was randomly determined.

\section{RESULTS}
In this section, we first collected baseline demographic information from all participants. Additionally, all participants with ASD were evaluated using the Autism Diagnostic Observation Schedule (ADOS\cite{lord2000autism}) and the Adaptive Behavior Assessment System, Second Edition (ABAS-II\cite{oakland2011adaptive,oakland2008adaptive}) during the baseline assessment. The baseline demographic information was analyzed to ensure that there were no statistically significant differences in the current levels of ASD symptoms among the participants in different groups.

\begin{table*}[]
\caption{Comparison in clinical baseline data among the three groups}
\centering
\begin{threeparttable}
\label{baseline}

\renewcommand{\arraystretch}{1.3}
\begin{tabular}{llllll}
\hline
                          & Control group & Avatar Uncustomized Group & Avatar Customized Group & F/Z/$\mathcal{X}^2$  & P     \\ \hline
Age                       & 3.71(0.75)    & 6.17(1.26)                & 4.24(1.80)              & $11.30^{\star}$ & 0.004 \\
Gender                    &               &                           &                         & $4.36^{\dagger}$   & 0.113 \\
$~~~~$male                      & 8(100\%)      & 6(75.00\%)                & 8(100\%)                &        &       \\
$~~~~$female                    & 0(0.00\%)     & 2(25.00\%)                & 0(0.00\%)               &        &       \\
ADOS                      &               &                           &                         &        &       \\
$~~~~$Css & 7.88(1.81)    & 7.50(1.07)                & 8.00(1.60)              & 0.23   & 0.794 \\
FSIQ                      & 75.63(12.85)  & 85.88(20.99)              & 82.25(18.95)            & 0.67   & 0.521 \\
ABAS-II                      &               &                           &                         &        &       \\
$~~~~$Overall adaptive skills   & 72.63(14.05)  & 86.88(22.21)              & 81.88(12.10)            & 1.50   & 0.246 \\
$~~~~$Conceptual skills         & 75.88(15.18)  & 85.25(19.60)              & 84.00(13.69)            & 0.78   & 0.473 \\
$~~~~$Social skills             & 69.00(18.99)  & 81.00(20.41)              & 81.50(13.15)            & 1.27   & 0.303 \\
$~~~~$Practical skills          & 71.88(15.34)  & 92.88(23.01)              & 82.25(12.08)            & 2.91   & 0.077 \\ \hline
\end{tabular}
    \begin{tablenotes}
       \footnotesize
       \item[1] ADOS-Css: Autism Diagnostic Observation Schedule-Calibrated severity score; ABAS-II: Adaptive Behavior Assessment System Version II; FSIQ: full scale intelligence quotient
       \item[2] $^{\dagger}$ Chi-square test; $^{\star}$ Kruskal-Wallis test; the others were One-Way Analysis of Variance
     \end{tablenotes}
    \end{threeparttable}
  \label{tb:1}%

\end{table*}

\begin{table*}[!t]
\caption{Parents' feedback from pre-test and post-test}
\label{questionnaire}
\centering

\resizebox{\linewidth}{!}{
\renewcommand{\arraystretch}{1.5}
\begin{tabular}{ p{3cm} p{1cm} p{1cm} p{1cm} p{1cm} p{1cm} p{1cm} p{1cm} p{1cm} p{1cm} p{1cm} p{1cm} p{1cm} }
\cline{1-13}
\multirow{3}{*}{Question} & \multicolumn{6}{c} {Pre-test} & \multicolumn{6}{c} {Post-test} \\
\cmidrule(r){2-7} \cmidrule(r){8-13}
 & \multicolumn{2}{c}{CG(n=8)} & \multicolumn{2}{c}{AUG(n=8)} & \multicolumn{2}{c}{ACG(n=8)} & \multicolumn{2}{c}{CG(n=8)} & \multicolumn{2}{c}{AUG(n=8)} & \multicolumn{2}{c}{ACG(n=8)} \\
\cmidrule(r){2-3}  \cmidrule(r){4-5}  \cmidrule(r){6-7} \cmidrule(r){8-9}   \cmidrule(r){10-11}  \cmidrule(r){12-13}
 & Mean & SD & Mean & SD & Mean & SD & Mean & SD & Mean& SD & Mean & SD \\
\hline
How often do kids notice your gaze? & 3.00 & 1.20 & 3.25 & 1.30 & 2.63 & 1.19 & 3.25 & 1.16 & 4.25 & 0.83 & 4.25& 0.71\\
\cline{1-13} 
Kids like makes eye contact with you? & 4.13 & 0.83 & 3.25 & 0.83 & 2.88 & 1.13 & 4.13 & 0.35 & 4.38 & 0.70 & 4.63 & 0.92\\
\cline{1-13} 
How often does your kid notice your pointing? & 3.63 & 0.92 & 4.25 & 1.20 & 4.25 & 1.40 & 3.88 & 0.99 & 4.50 & 0.71 & 5.38 & 0.74\\
\cline{1-13} 
How often does your kid point? & 3.25 & 1.16 & 4.75 & 0.97 & 4.00 & 0.53 & 3.50 & 0.76 & 4.88 & 0.78 & 5.50 & 0.93\\
\cline{1-13} 
How often does your kid respond to their name? & 3.25 & 1.49 & 4.13 & 1.17 & 3.63 & 0.92 & 3.63 & 0.74 & 4.50 & 1.00 & 4.50 & 0.76\\
\cline{1-13} 
How often does your kid respond to sounds? & 3.63 & 1.19 & 4.50 & 1.12 & 3.75 & 1.83 & 3.75 & 0.71 & 5.00 & 0.71 & 5.13 & 0.83\\
\cline{1-13} 
Kids like hide-and-seek games? & 3.13 & 1.36 & 4.75 & 1.30 & 4.50& 1.41 & 3.88 & 1.36 & 5.50 & 0.71 & 5.25 & 0.89\\
\cline{1-13} 
Kids like respond to verbal cues in hide and seek? & 3.50 & 1.41 & 4.50 & 1.22 & 5.13 & 0.83 & 3.50 & 1.51 & 5.13 & 0.78 & 5.63 & 0.74\\
\cline{1-13}
Kids like respond to body cues in hide and seek? & 3.75 & 1.19 & 4.25 & 1.20 & 4.63 & 1.60 & 4.25 & 1.28 & 5.25 & 0.83 & 5.75 & 0.46\\
\cline{1-13}
Kids like respond to others' cues in hide and seek? & 3.38 & 1.18 & 3.50 & 1.12 & 5.00 & 1.07 & 3.88 & 0.64 & 5.13 & 0.78 & 5.75 & 0.71\\
\cline{1-13}
\end{tabular}}
   \begin{threeparttable}
    \begin{tablenotes}
       \footnotesize
       \item[1] CG: Control Group; AUG: Avatar Uncustomized Group; ACG: Avatar Customized Group 
     \end{tablenotes}
    \end{threeparttable}
\end{table*}

\begin{table*}[!t]
\caption{Mixed linear regression model in comparing the difference in questionnaire over time between the control group and the avatar uncustomized group}
\label{question1}
\centering
\resizebox{\textwidth}{!}{
\renewcommand{\arraystretch}{1.5}
\begin{tabular}{ c c c c c c c c c}
\hline
Dependent variables & \multicolumn{2}{c}{Intercept} & \multicolumn{2}{c}{Group effect} & \multicolumn{2}{c}{Time effect} & \multicolumn{2}{c}{Time-by-group effect}\\
\cmidrule(r){2-3} \cmidrule(r){4-5} \cmidrule(r){6-7} \cmidrule(r){8-9}
 & B (SE) & P value & B (SE) & P value & B (SE) & P value & B (SE) & P value \\
\hline 
Questionnaire Points & & & & & & & &\\
Baseline & 28.47(4.87) & $<0.001$ & 5.41(3.08) & 0.095& & & & \\
T0-T4 & & & & & -0.34(0.54) & 0.53 & 1.09(0.34) & \textbf{0.006}\\

\hline
\end{tabular}}
\end{table*}

\begin{table*}[!t]
\caption{Mixed linear regression model in comparing the difference in questionnaire over time between the control group and the avatar customized group}
\label{question2}
\centering
\resizebox{\textwidth}{!}{
\renewcommand{\arraystretch}{1.5}
\begin{tabular}{ c c c c c c c c c}
\hline
Dependent variables & \multicolumn{2}{c}{Intercept} & \multicolumn{2}{c}{Group effect} & \multicolumn{2}{c}{Time effect} & \multicolumn{2}{c}{Time-by-group effect}\\
\cmidrule(r){2-3} \cmidrule(r){4-5} \cmidrule(r){6-7} \cmidrule(r){8-9}
 & B (SE) & P value & B (SE) & P value & B (SE) & P value & B (SE) & P value \\
\hline 
Questionnaire Points & & & & & & & &\\
Baseline & 32.04(3.28) & $<0.001$ & 1.83(1.47) & 0.228& & & & \\
T0-T4 & & & & & -0.30(0.38) & 0.441 & 1.05(0.168) & $<\textbf{0.001}$\\

\hline
\end{tabular}}
\end{table*}

\begin{table*}[!t]
\caption{Mixed linear regression model in comparing the difference in questionnaire over time between the avatar uncustomized group and the avatar customized group}
\label{question3}
\centering
\resizebox{\textwidth}{!}{
\renewcommand{\arraystretch}{1.5}
\begin{tabular}{ c c c c c c c c c}
\hline
Dependent variables & \multicolumn{2}{c}{Intercept} & \multicolumn{2}{c}{Group effect} & \multicolumn{2}{c}{Time effect} & \multicolumn{2}{c}{Time-by-group effect}\\
\cmidrule(r){2-3} \cmidrule(r){4-5} \cmidrule(r){6-7} \cmidrule(r){8-9}
 & B (SE) & P value & B (SE) & P value & B (SE) & P value & B (SE) & P value \\
\hline 
Questionnaire Points & & & & & & & &\\
Baseline & 42.78(8.18) & $<0.001$ & -1.75(3.21) & 0.591& & & & \\
T0-T4 & & & & & -0.16(1.04) & 0.883 & 1.00(0.41) & \textbf{0.027}\\

\hline
\end{tabular}}
\end{table*}

Next, we analyzed the subjective questionnaires completed by the parents of the participants. The results of the questionnaire analysis are presented to demonstrate the positive effects of using our HSVRS as an auxiliary intervention therapy.

Furthermore, we present the objective results from two aspects: the player's game performance and the player's gaze fixation. By comparing these results, we conclude that: i) our HSVRS is effective in improving children's gaze fixation ability, and ii) the use of deep learning-based face and voice manipulation technology can enhance the effectiveness of the intervention.

    


    

    

\begin{figure}[htbp]
    \subfigure[Time to Complete]{
    \begin{minipage}{1.0\linewidth}
    \includegraphics[width=3.4in]{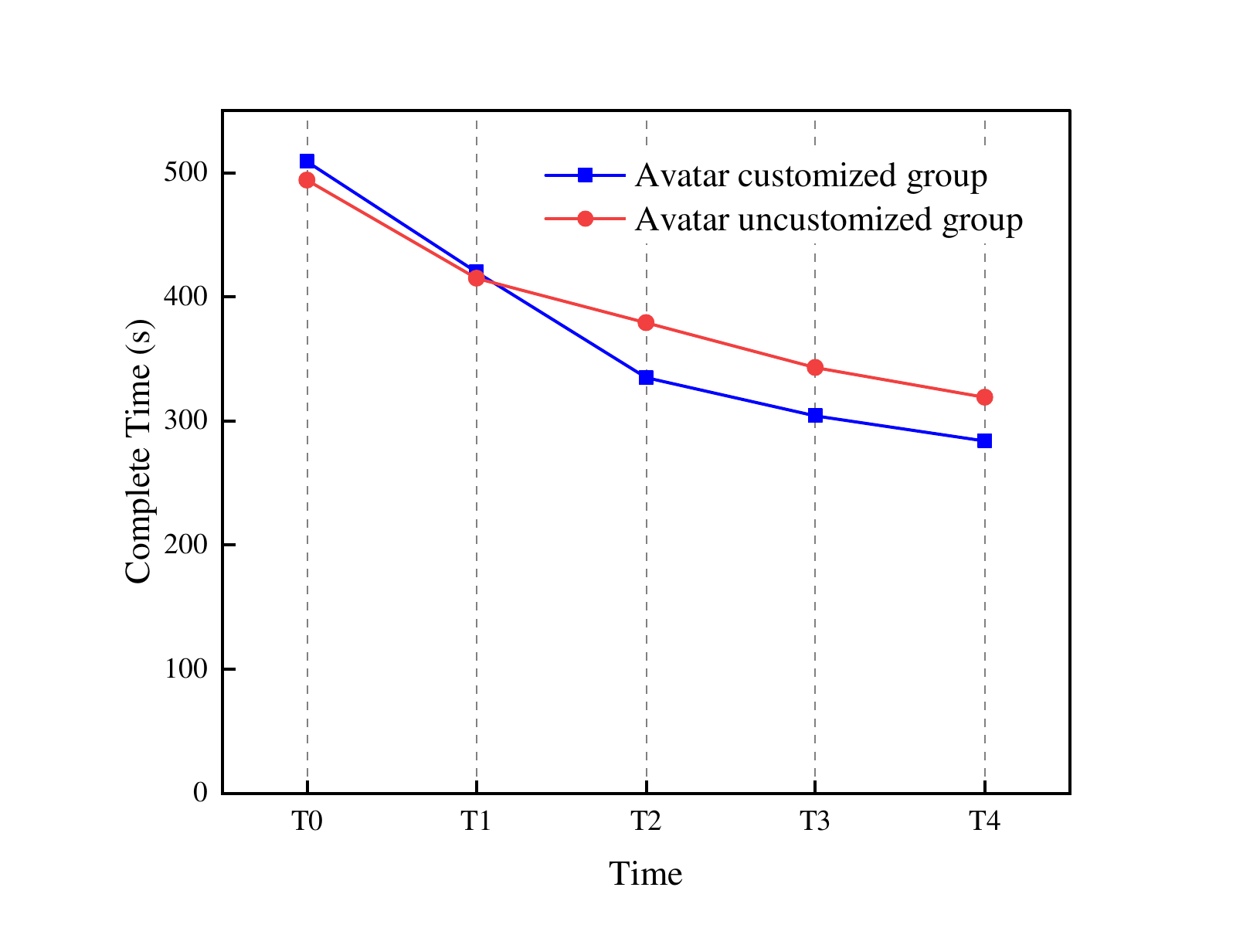}
    \label{completion time}
    \end{minipage}
    }

    \subfigure[Average Response Time]{
    \begin{minipage}{1.0\linewidth}
    \includegraphics[width=3.4in]{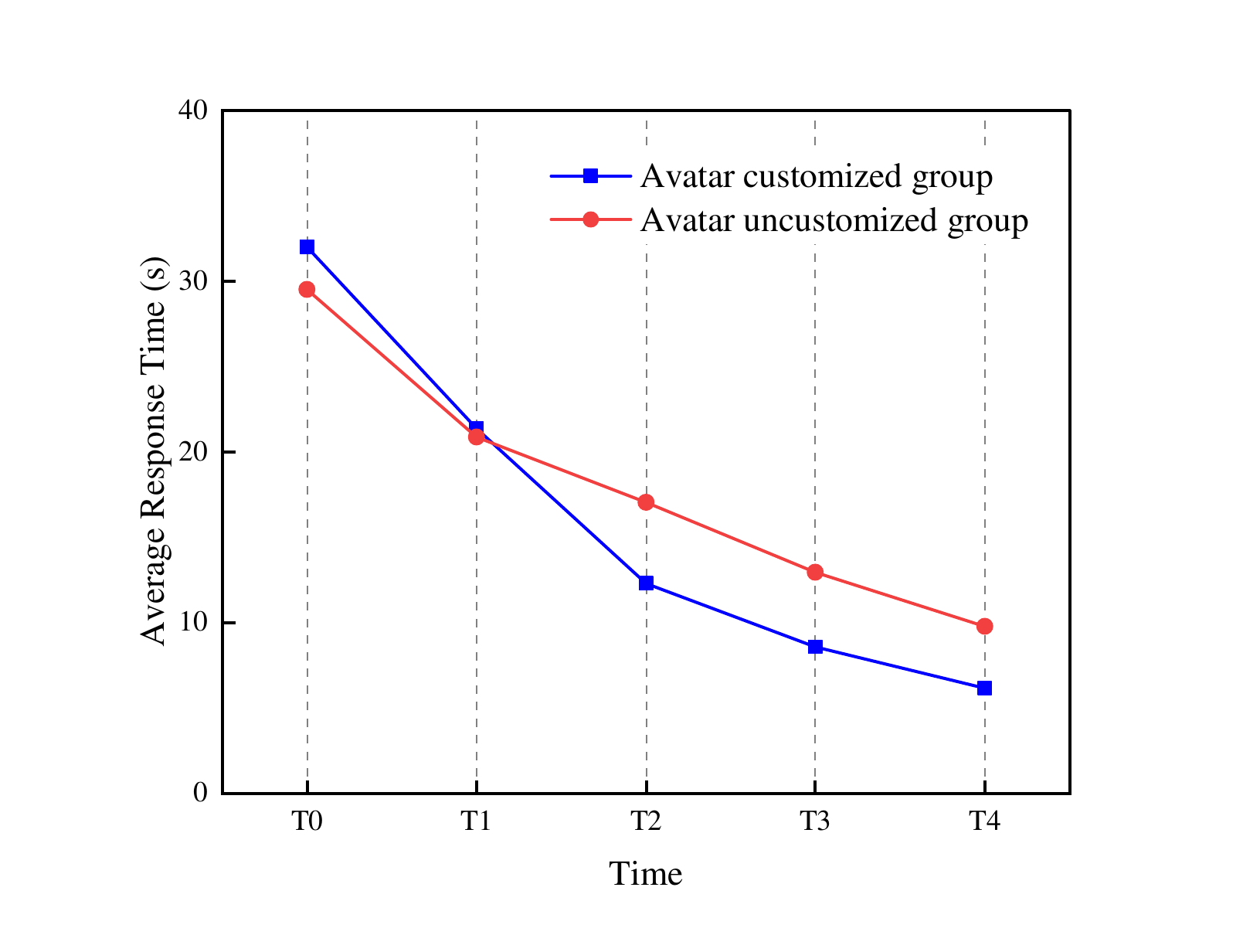}
    \label{Average Response Time}
    \end{minipage}
    }

    \subfigure[Game Score]{
    \begin{minipage}{1.0\linewidth}
    \includegraphics[width=3.4in]{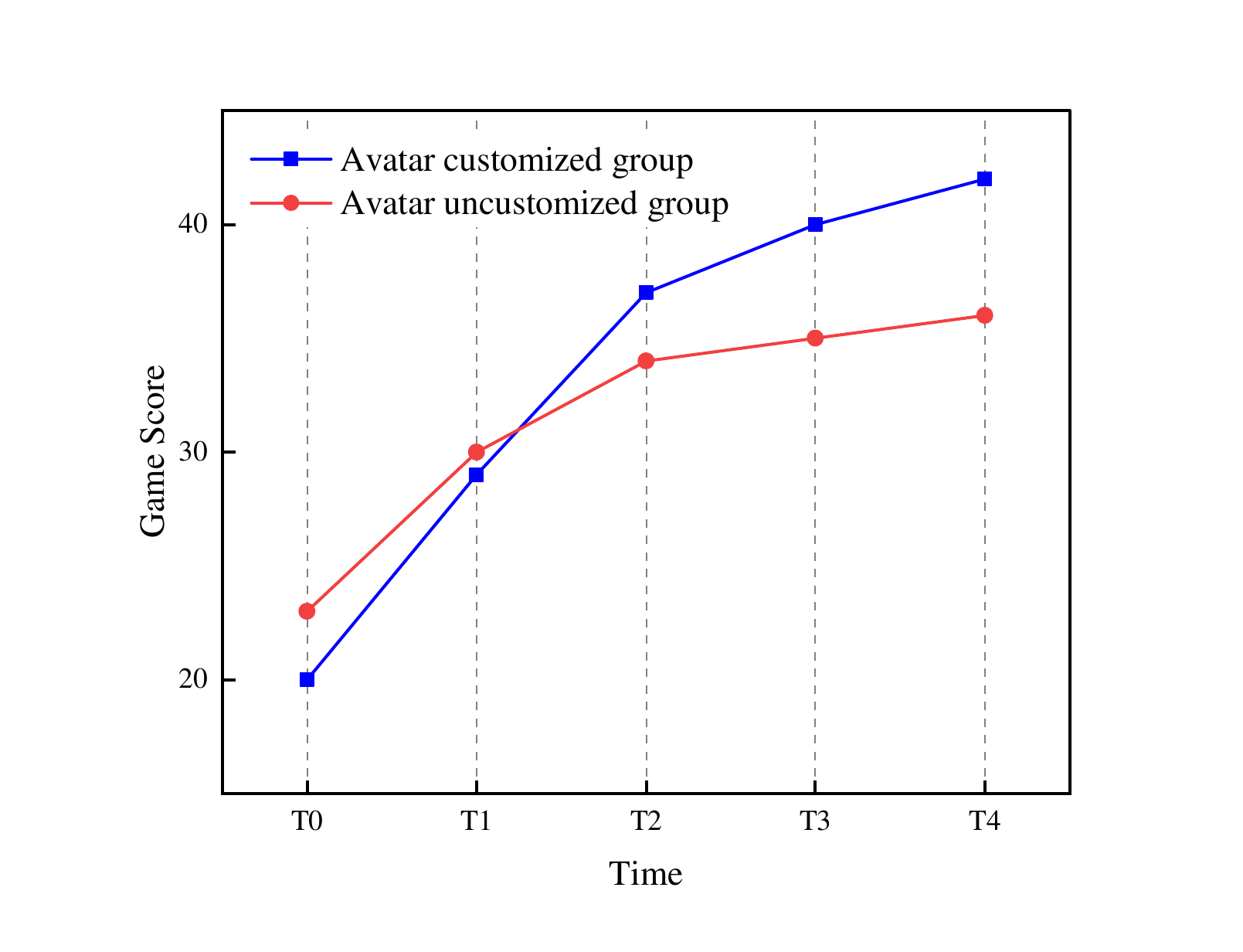}
    \label{Game Score}
    \end{minipage}
    }    
    \caption{Line charts of the changes in the three game performance metrics over the intervention period.}
    \label{threeFigs}

\end{figure}

In the analysis of the experimental results, we utilized SPSS 26.0 software. Continuous data were presented as means and standard deviations (SD). A significance level of $P<0.05$ was considered statistically significant.
\subsection{Comparison in clinical baseline data}
There is a significant difference in age among the three groups ($Z=11.30, P=0.004$), while no statistical difference in sex ratio ($\mathcal{X}^2=0.36, P=0.113$). In order to reduce the difference in language levels and age, we applied standardized ADOS-Calibrated Severity Scores (Css) to evaluate the severe symptoms of autism. The results showed no statistical difference in ADOS-Css among the three groups ($F= 0.23, P= 0.794$), suggesting no significant difference in the severe symptoms of autism among the three groups. Similarly, standardized scores were applied to calculate age-normalized Intelligence quotient (IQ) and adaptive skills. There are no statistical differences in IQ and adaptive skills among the three groups (all $P>0.05$), which indicated no significant difference in cognitive functioning and adaptive functioning among the three groups.

\subsection{Analysis of subjective questionnaire}
Before and after the experimental period, we collected parental feedback from the participants using a subjective questionnaire consisting of ten questions. The questions, along with the parents' responses from the pre-test and post-test, are presented in Table \ref{questionnaire}. A six-point Likert scale was used for all the questions. The questionnaire was divided into two parts, with the first six questions relating to a life scenario and the last four questions pertaining to the context of a hide-and-seek game. These questions assessed the child's ability to make eye contact, point with their finger, and engage in visual and auditory reasoning, all of which are skills addressed in our game. The total score of the questionnaire provided feedback on the child's overall ability.
To compare the baseline data between the three groups, we conducted student t-tests on the total questionnaire scores. The results indicated no significant differences in the total questionnaire scores between the three groups (all $P>0.05$).

We then divide the collected questionnaires into three groups: the control group, the avatar uncustomized group and the avatar customized group. Considering the repeated collection of the questionnaire, we conduct three mixed linear regression models with two components: 1) a random-effect parameter for each participant to account for unobserved sources of heterogeneity among participants, and 2) two fixed-effect parameters corresponding to the group and time. 

Table \ref{question1} compares the difference in questionnaire points over time between the control group and the avatar uncustomized group. The score increased significantly between T4 and T0 in both groups ($P<0.001$). Moreover, there are significant increment differences in questionnaire scores between the control group and the avatar uncustomized group. The increments of questionnaire scores were significantly larger in the avatar uncustomized group than in the control group. ( $B=1.09;P=0.006$). 

The change in questionnaire points over time between the control group and the avatar customized group is compared in Table \ref{question2} between the two groups. Between T4 and T0, both groups' scores rose significantly ($P< 0.001$). Additionally, there are significant increases in questionnaire score discrepancies between the avatar customized group and the group that received no supplemental intervention. The increments in questionnaire scores in the avatar customized group were substantially more prominent than in the group that did not receive any auxiliary interventions ($B=1.05; P<0.001$).

Table \ref{question3} compares the difference in questionnaire points over time between the avatar uncustomized group and the avatar customized group. The score increased significantly between T4 and T0 in both groups ($P<0.001$). And there are significant increment differences in questionnaire scores between the avatar uncustomized group and the avatar customized group. The increments of questionnaire scores were significantly more prominent in the avatar customized group than in the avatar uncustomized group ( $B= 1.00; P=0.027$).

In summary, the VR auxiliary intervention group was more effective than the control group. Moreover, the best intervention results were achieved if avatars within the VR game were customized using face and voice deepfake manipulation techniques.

\begin{table}[]
\caption{List of performance metrics}
\label{performance metric}
\centering
\begin{tabular}{@{}ll@{}}
\toprule
Performance Metric & Description \\ \midrule
\begin{tabular}[c]{@{}l@{}}completion time \\ (seconds)\end{tabular} &
  \begin{tabular}[c]{@{}l@{}}The total time it takes for a participant\\ to complete the game. In each game, the\\ parent avatar hides in nine different random\\ places. When the participant finds all nine \\ places, the game is terminated.\end{tabular} \\ \midrule
\begin{tabular}[c]{@{}l@{}}Average Response\\ Time (seconds)\end{tabular} &
  \begin{tabular}[c]{@{}l@{}}Response time is computed between the time\\ when the parent avatar hides properly and\\ the time when the participant finds the \\ hidden parent avatar.\end{tabular} \\ \midrule
\begin{tabular}[c]{@{}l@{}}Game Score \\ (points)\end{tabular} &
  \begin{tabular}[c]{@{}l@{}}In each round of the game, five points are\\ received if the participant finds the parent\\ avatar without any prompts. The game score \\ decreases as the intensity of the prompt \\ participant get increases. The maximum \\ possible score is 45.\end{tabular} \\ \bottomrule
\end{tabular}
\end{table}

\subsection{Analysis of participants' game performance}

Game performance is measured using game scores, completion times, and average response times. Table \ref{performance metric} lists the metrics together with a description of each metric. Fig. \ref{threeFigs} shows the line charts of the changes in the three game performance metrics over the intervention period. Participants' game performance, including completion time, average response time and game score, increase significantly between T4 and T0 in both groups. Specifically, children in the avatar customized group took less time to complete the game, their average response time decreased, and they received higher game scores. 
For the completion time metric, the avatar uncustomized group and the avatar customized group had  similar levels at T0, where they took about 500 seconds to complete the game. However, from Fig. \ref{threeFigs}, we find that children from the avatar customized group exhibited more significant improvement since the curve from the avatar customized group had a higher slope. 
For the average response time metric, the avatar uncustomized group and the avatar customized group showed similar levels at T0, where it takes roughly 32 seconds to discover the parent avatar in each round. Then, Fig. \ref{threeFigs} shows that the average response time of the avatar-customized group reduces more quickly, suggesting that the children in this group received a better boost.
For the game score metric, the average game score of the avatar customized group was lower than that of the uncustomized group at T0. Nevertheless, at T4, the average game score of the avatar customized group was higher than that of the uncustomized group. It indicates that children from the avatar customized group exhibited more remarkable improvement. 
\begin{table*}[!t]
\caption{Comparison in completion time, average response time and game socre metric between the avatar uncustomized group and the avatar customized group in baseline (T0) assessment}
\label{baselineComparison}
\centering
\renewcommand{\arraystretch}{1.5}
\begin{tabular}{ p{3.5cm} p{3.5cm} p{3.5cm} p{2cm} p{2cm}}
\hline
Dependent variables & avatar uncustomized group & avatar customized group & t value & P value\\
\hline
Game Performance & & & & \\
$~~~~$completion time & 505.50 (120.26) & 509.38 (69.86) & 0.079 & 0.938\\
$~~~~$Average Response Time & 30.94 (14.56) & 32.01 (8.70) & 0.180 & 0.860\\
$~~~~$Game Score & 22.75 (7.48)& 20.25 (7.72) & -0.658 & 0.521\\
\hline
\end{tabular}
\end{table*}

\begin{table*}[!t]
\caption{Mixed linear regression model in comparing the difference in game performance over time between the avatar uncustomized group and the avatar customized group}
\label{mixed linear model}
\centering
\resizebox{\textwidth}{!}{
\renewcommand{\arraystretch}{1.5}
\begin{tabular}{ c c c c c c c c c}
\hline
Dependent variables & \multicolumn{2}{c}{Intercept} & \multicolumn{2}{c}{Group effect} & \multicolumn{2}{c}{Time effect} & \multicolumn{2}{c}{Time-by-group effect}\\
\cmidrule(r){2-3} \cmidrule(r){4-5} \cmidrule(r){6-7} \cmidrule(r){8-9}
 & B (SE) & P value & B (SE) & P value & B (SE) & P value & B (SE) & P value \\
\hline 
completion time & & & & & & & &\\
Baseline & 515.99(23.13) & $<0.001$ & -42.02(32.71) & 0.467& & & & \\
T0-T4 & & & & & -42.03(5.20) & $<0.001$ & -14.49(7.35) & 0.053\\
Average response time & & & & & & & &\\
Baseline & 32.26(2.88) & $<0.001$ & 3.18(4.07) & 0.440& & & & \\
T0-T4 & & & & & -4.74(0.56) & $<0.001$ & -1.71(0.80) & $\mathbf{0.036}$\\
Game Score & & & & & & & &\\
Baseline & 22.01(2.62) & $<0.001$ & -4.60(3.71) & 0.221& & & & \\
T0-T4 & & & & & 3.01(0.56) & $<0.001$ & 2.33(0.79) & $\mathbf{0.005}$\\
\hline
\end{tabular}}
\end{table*}

Table \ref{baselineComparison} shows the baseline assessment of the three game performance metrics. No statistical differences were found in completion time, average response time and game score between the avatar uncustomized group and the avatar customized group (all $P>0.05$). It demonstrates that there is no significant difference between the avatar uncustomized group and the avatar customized group when children get started to play hide-and-seek in HRVRS.

For more detailed and convincing data analysis, as well as considering the repeated measures design of our study, a mixed linear regression model was applied to analyze outcome variables by modeling all five measurements (T0, T1, T2, T3 and T4).

Completion time, average response time and the game score increased significantly between T4 and T0 in both groups (all $P<0.05$). After one week's auxiliary intervention, the increments in average response time and game score metrics were significantly larger in the avatar customized group than in the avatar uncustomized group ($B$=-1.71,2.33, $P$=0.036,0.005), whereas no significant differences in the increment scores were found in completion time metric($P$=0.053). Thus, in summary, participants from the avatar customized group were able to achieve better intervention results.

\subsection{Gaze Fixation Analysis}
To better understand the distribution of the participants' fixation during the hide-and-seek game, we predefined three ROIs (regions of interest), namely face region, body region and background region. To obtain fixation metrics for these ROIs, we design an algorithm based on the 7Invensun Unity development package. For details, we record the fixation points of the gaze when the participants are playing the game, and then we divide the total fixation points into three categories based on their location. In the data analysis, we normalize these three metrics. The normalized results represent the fraction of fixation points from each ROI. From Table \ref{gazeFixation}, we can observe that there is a statistically significant difference in face fixation proportion and background fixation proportions between the avatar uncustomized group and the avatar customized group (all $P<0.001$). Meanwhile, there is no statistical difference in body fixation proportion between the avatar uncustomized group and the avatar customized group ($P=0.306$). It demonstrate that using face and voice manipulation techniques to customize the characters in games made participants more likely to look at the facial regions of the avatars.

\begin{table}
\caption{Results for gaze fixations}
\centering
\renewcommand{\arraystretch}{1.5}
\begin{tabular}{p{2cm} p{1.8cm} p{1.5cm} p{1cm} p{1cm}}
\cmidrule(r){1-5}
Metrics & avatar uncustomized group & avatar customized group & & \\
\cmidrule(r){2-5}
& Mean(SD) & Mean(SD) & t value & P value \\
\cmidrule(r){1-5}
Face Fixation Propotion & 12.52(4.33) & 23.23(8.14) & 7.34 & $<\textbf{0.001}$ \\
Body Fixation Propotion & 22.68(8.21) & 21.09(5.36) & -1.03 & 0.306 \\
Background Fixation Propotion & 64.79(11.30) & 55.69(11.74) & -3.536 & $<\textbf{0.001}$\\
\cmidrule(r){1-5}
\end{tabular}

\label{gazeFixation}
\end{table}

\section{DISCUSSION AND CONCLUSION}

In this study, we develop a novel VR system called HSVRS to target gaze fixation skills in autistic children. The system incorporates various components to enhance the intervention's effectiveness. The Avatar Controller enables avatar customization using computer vision and voice conversion techniques. The Prompt Controller accommodates diverse learning abilities by providing sequential prompts. The Game Module coordinates the game operation, while the Reinforcement Module provides rewards and positive feedback. The Gaze Module utilizes eye tracking for real-time analysis, and the Data Logger records all game and gaze data for analysis.

The pilot study aims to evaluate the efficacy of hide-and-seek games in VR headsets, specifically focusing on improving gaze fixation in children with ASD. We hypothesize that using HSVRS would lead to improved gaze fixation and that the avatar customized group would exhibit larger increments than the uncustomized avatar group. The participants are divided into three groups, and repeated measurements are conducted over a one-week period. The experimental results demonstrate the potential of the HSVRS system to improve gaze fixation in children with ASD. Both groups show improvements in game performance, as evidenced by shorter completion times, shorter average response times, and higher game scores. Furthermore, the mixed linear model analysis reveals significant and consistent inter-group differences. After one week of intervention, the avatar customized group exhibits significantly larger increments in average response time and game score compared to the avatar uncustomized group. These findings support the hypothesis that in-game avatar customization enhances intervention effectiveness. The gaze data analysis reveals that children from the avatar customized group focus more on the face region of the parent avatar, indicating the positive impact of face manipulation design in promoting eye contact.

While the results are encouraging, it is important to acknowledge the limitations of this study and identify areas for future investigation. The sample size is relatively small, and a larger longitudinal study would provide more robust assessments. Additionally, future studies should include pre- and post-evaluations for the control group to further evaluate the impact of HSVRS on gaze fixation. Lastly, incorporating rich facial expressions for avatars during gameplay could be a potential avenue for further exploration. Overall, the pilot study demonstrates the potential benefits of using HSVRS for improving gaze fixation in children with ASD, providing a foundation for future research in this field.

\section*{Acknowledgment}

The authors would like to thank...

\ifCLASSOPTIONcaptionsoff
  \newpage
\fi



\bibliographystyle{IEEEtran}

\begin{thebibliography}{10}
\providecommand{\url}[1]{#1}
\csname url@samestyle\endcsname
\providecommand{\newblock}{\relax}
\providecommand{\bibinfo}[2]{#2}
\providecommand{\BIBentrySTDinterwordspacing}{\spaceskip=0pt\relax}
\providecommand{\BIBentryALTinterwordstretchfactor}{4}
\providecommand{\BIBentryALTinterwordspacing}{\spaceskip=\fontdimen2\font plus
\BIBentryALTinterwordstretchfactor\fontdimen3\font minus
  \fontdimen4\font\relax}
\providecommand{\BIBforeignlanguage}[2]{{%
\expandafter\ifx\csname l@#1\endcsname\relax
\typeout{** WARNING: IEEEtran.bst: No hyphenation pattern has been}%
\typeout{** loaded for the language `#1'. Using the pattern for}%
\typeout{** the default language instead.}%
\else
\language=\csname l@#1\endcsname
\fi
#2}}
\providecommand{\BIBdecl}{\relax}
\BIBdecl

\bibitem{lord2018autism}
C.~Lord, M.~Elsabbagh, G.~Baird, and J.~Veenstra-Vanderweele, ``Autism spectrum
  disorder,'' \emph{The lancet}, vol. 392, no. 10146, pp. 508--520, 2018.

\bibitem{american2013dsm}
A.~P. Association, A.~P. Association \emph{et~al.}, ``Dsm-5 task force
  diagnostic and statistical manual of mental disorders: Dsm-5. washington,''
  \emph{DC: American Psychiatric Association}, 2013.

\bibitem{maenner2021prevalence}
M.~J. Maenner, K.~A. Shaw, A.~V. Bakian, D.~A. Bilder, M.~S. Durkin, A.~Esler,
  S.~M. Furnier, L.~Hallas, J.~Hall-Lande, A.~Hudson \emph{et~al.},
  ``Prevalence and characteristics of autism spectrum disorder among children
  aged 8 years—autism and developmental disabilities monitoring network, 11
  sites, united states, 2018,'' \emph{MMWR Surveillance Summaries}, vol.~70,
  no.~11, p.~1, 2021.

\bibitem{webb2014motivation}
S.~J. Webb, E.~J. Jones, J.~Kelly, and G.~Dawson, ``The motivation for very
  early intervention for infants at high risk for autism spectrum disorders,''
  \emph{International journal of speech-language pathology}, vol.~16, no.~1,
  pp. 36--42, 2014.

\bibitem{chevallier2015measuring}
C.~Chevallier, J.~Parish-Morris, A.~McVey, K.~M. Rump, N.~J. Sasson, J.~D.
  Herrington, and R.~T. Schultz, ``Measuring social attention and motivation in
  autism spectrum disorder using eye-tracking: Stimulus type matters,''
  \emph{Autism Research}, vol.~8, no.~5, pp. 620--628, 2015.

\bibitem{jones2013attention}
W.~Jones and A.~Klin, ``Attention to eyes is present but in decline in
  2--6-month-old infants later diagnosed with autism,'' \emph{Nature}, vol.
  504, no. 7480, pp. 427--431, 2013.

\bibitem{frazier2017meta}
T.~W. Frazier, M.~Strauss, E.~W. Klingemier, E.~E. Zetzer, A.~Y. Hardan,
  C.~Eng, and E.~A. Youngstrom, ``A meta-analysis of gaze differences to social
  and nonsocial information between individuals with and without autism,''
  \emph{Journal of the American Academy of Child \& Adolescent Psychiatry},
  vol.~56, no.~7, pp. 546--555, 2017.

\bibitem{towle2020earlier}
P.~O. Towle, P.~A. Patrick, T.~Ridgard, S.~Pham, and J.~Marrus, ``Is earlier
  better? the relationship between age when starting early intervention and
  outcomes for children with autism spectrum disorder: a selective review,''
  \emph{Autism Research and Treatment}, vol. 2020, 2020.

\bibitem{smith2008roles}
M.~M. Smith and I.~Connolly, ``Roles of aided communication: Perspectives of
  adults who use aac,'' \emph{Disability and rehabilitation: Assistive
  technology}, vol.~3, no.~5, pp. 260--273, 2008.

\bibitem{parsons2011state}
S.~Parsons and S.~Cobb, ``State-of-the-art of virtual reality technologies for
  children on the autism spectrum,'' \emph{European Journal of Special Needs
  Education}, vol.~26, no.~3, pp. 355--366, 2011.

\bibitem{little2016gaze}
G.~Little, L.~Bonnar, S.~Kelly, K.~S. Lohan, and G.~Rajendran, ``Gaze
  contingent joint attention with an avatar in children with and without asd,''
  in \emph{2016 joint IEEE international conference on development and learning
  and epigenetic robotics (ICDL-EpiRob)}.\hskip 1em plus 0.5em minus
  0.4em\relax IEEE, 2016, pp. 15--20.

\bibitem{caruana2018joint}
N.~Caruana, H.~Stieglitz~Ham, J.~Brock, A.~Woolgar, N.~Kloth, R.~Palermo, and
  G.~McArthur, ``Joint attention difficulties in autistic adults: an
  interactive eye-tracking study,'' \emph{Autism}, vol.~22, no.~4, pp.
  502--512, 2018.

\bibitem{courgeon2014joint}
M.~Courgeon, G.~Rautureau, J.-C. Martin, and O.~Grynszpan, ``Joint attention
  simulation using eye-tracking and virtual humans,'' \emph{IEEE Transactions
  on Affective Computing}, vol.~5, no.~3, pp. 238--250, 2014.

\bibitem{yaneva2020detecting}
V.~Yaneva, S.~Eraslan, Y.~Yesilada, R.~Mitkov \emph{et~al.}, ``Detecting
  high-functioning autism in adults using eye tracking and machine learning,''
  \emph{IEEE Transactions on Neural Systems and Rehabilitation Engineering},
  vol.~28, no.~6, pp. 1254--1261, 2020.

\bibitem{cooper2007applied}
J.~O. Cooper, T.~E. Heron, W.~L. Heward \emph{et~al.}, ``Applied behavior
  analysis,'' 2007.

\bibitem{dawson2010randomized}
G.~Dawson, S.~Rogers, J.~Munson, M.~Smith, J.~Winter, J.~Greenson,
  A.~Donaldson, and J.~Varley, ``Randomized, controlled trial of an
  intervention for toddlers with autism: the early start denver model,''
  \emph{Pediatrics}, vol. 125, no.~1, pp. e17--e23, 2010.

\bibitem{bekteshi2020eye}
S.~Bekteshi, M.~Konings, I.~Vanmechelen, J.~Deklerck, E.~Ortibus, J.-M. Aerts,
  H.~Hallez, P.~Karlsson, B.~Dan, and E.~Monbaliu, ``Eye gaze gaming
  intervention in children with dyskinetic cerebral palsy: a pilot study of
  task performance and its relation with dystonia and choreoathetosis,''
  \emph{Developmental neurorehabilitation}, vol.~23, no.~8, pp. 548--556, 2020.

\bibitem{carlon2013review}
S.~Carlon, M.~Carter, and J.~Stephenson, ``A review of declared factors
  identified by parents of children with autism spectrum disorders (asd) in
  making intervention decisions,'' \emph{Research in Autism Spectrum
  Disorders}, vol.~7, no.~2, pp. 369--381, 2013.

\bibitem{berton2020eye}
F.~Berton, L.~Hoyet, A.-H. Olivier, J.~Bruneau, O.~Le~Meur, and J.~Pettr{\'e},
  ``Eye-gaze activity in crowds: impact of virtual reality and density,'' in
  \emph{2020 IEEE Conference on Virtual Reality and 3D User Interfaces
  (VR)}.\hskip 1em plus 0.5em minus 0.4em\relax IEEE, 2020, pp. 322--331.

\bibitem{berton2019studying}
F.~Berton, A.-H. Olivier, J.~Bruneau, L.~Hoyet, and J.~Pettr{\'e}, ``Studying
  gaze behaviour during collision avoidance with a virtual walker: Influence of
  the virtual reality setup,'' in \emph{2019 IEEE Conference on Virtual Reality
  and 3D User Interfaces (VR)}.\hskip 1em plus 0.5em minus 0.4em\relax IEEE,
  2019, pp. 717--725.

\bibitem{guo2020towards}
J.~Guo, X.~Zhu, Y.~Yang, F.~Yang, Z.~Lei, and S.~Z. Li, ``Towards fast,
  accurate and stable 3d dense face alignment,'' in \emph{Proceedings of the
  European Conference on Computer Vision (ECCV)}, 2020, pp. 152--168.

\bibitem{zhu2016face}
X.~Zhu, Z.~Lei, X.~Liu, H.~Shi, and S.~Z. Li, ``Face alignment across large
  poses: A 3d solution,'' in \emph{2016 IEEE Conference on Computer Vision and
  Pattern Recognition (CVPR)}.\hskip 1em plus 0.5em minus 0.4em\relax IEEE
  Computer Society, 2016, pp. 146--155.

\bibitem{zhu2017face}
X.~Zhu, X.~Liu, Z.~Lei, and S.~Z. Li, ``Face alignment in full pose range: A 3d
  total solution,'' \emph{IEEE transactions on pattern analysis and machine
  intelligence}, vol.~41, no.~1, pp. 78--92, 2017.

\bibitem{vaswani2017attention}
A.~Vaswani, N.~Shazeer, N.~Parmar, J.~Uszkoreit, L.~Jones, A.~N. Gomez,
  {\L}.~Kaiser, and I.~Polosukhin, ``Attention is all you need,''
  \emph{Advances in neural information processing systems}, vol.~30, 2017.

\bibitem{zhang2022sig}
H.~Zhang, Z.~Cai, X.~Qin, and M.~Li, ``Sig-vc: A speaker information guided
  zero-shot voice conversion system for both human beings and machines,'' in
  \emph{ICASSP 2022-2022 IEEE International Conference on Acoustics, Speech and
  Signal Processing (ICASSP)}.\hskip 1em plus 0.5em minus 0.4em\relax IEEE,
  2022, pp. 6567--65\,571.

\bibitem{cai2020speaker}
Z.~Cai, C.~Zhang, and M.~Li, ``From speaker verification to multispeaker speech
  synthesis, deep transfer with feedback constraint,'' in \emph{Proc. of
  Interspeech}, 2020, pp. 3974--3978.

\bibitem{du2021optimizing}
H.~Du, X.~Tian, L.~Xie, and H.~Li, ``Optimizing voice conversion network with
  cycle consistency loss of speaker identity,'' in \emph{2021 IEEE Spoken
  Language Technology Workshop (SLT)}.\hskip 1em plus 0.5em minus 0.4em\relax
  IEEE, 2021, pp. 507--513.

\bibitem{libby2008comparison}
M.~E. Libby, J.~S. Weiss, S.~Bancroft, and W.~H. Ahearn, ``A comparison of
  most-to-least and least-to-most prompting on the acquisition of solitary play
  skills,'' \emph{Behavior analysis in practice}, vol.~1, no.~1, pp. 37--43,
  2008.

\bibitem{polick2012comparison}
A.~S. Polick, J.~E. Carr, and N.~M. Hanney, ``A comparison of general and
  descriptive praise in teaching intraverbal behavior to children with
  autism,'' \emph{Journal of Applied Behavior Analysis}, vol.~45, no.~3, pp.
  593--599, 2012.

\bibitem{finke2017effects}
E.~H. Finke, J.~M. Davis, M.~Benedict, L.~Goga, J.~Kelly, L.~Palumbo, T.~Peart,
  and S.~Waters, ``Effects of a least-to-most prompting procedure on
  multisymbol message production in children with autism spectrum disorder who
  use augmentative and alternative communication,'' \emph{American journal of
  speech-language pathology}, vol.~26, no.~1, pp. 81--98, 2017.

\bibitem{yanardaug2011effects}
M.~Yanardag, B.~Birkan, {\.I}.~Y{\i}lmaz, F.~K. Konukman,
  B.~A{\u{g}}bu{\u{g}}a, and L.~Lieberman, ``The effects of least-to-most
  prompting procedure in teaching basic tennis skills to children with
  autism,'' \emph{Kinesiology}, vol.~43, no.~1., pp. 44--55, 2011.

\bibitem{haxby2002human}
J.~V. Haxby, E.~A. Hoffman, and M.~I. Gobbini, ``Human neural systems for face
  recognition and social communication,'' \emph{Biological psychiatry},
  vol.~51, no.~1, pp. 59--67, 2002.

\bibitem{brodhead2014use}
M.~T. Brodhead, T.~S. Higbee, J.~S. Pollard, J.~S. Akers, and K.~R. Gerencser,
  ``The use of linked activity schedules to teach children with autism to play
  hide-and-seek,'' \emph{Journal of applied behavior analysis}, vol.~47, no.~3,
  pp. 645--650, 2014.

\bibitem{williams1981response}
J.~A. Williams, R.~L. Koegel, and A.~L. Egel, ``Response-reinforcer
  relationships and improved learning in autistic children,'' \emph{Journal of
  Applied Behavior Analysis}, vol.~14, no.~1, pp. 53--60, 1981.

\bibitem{charlop1998using}
M.~H. Charlop-Christy and L.~K. Haymes, ``Using objects of obsession as token
  reinforcers for children with autism,'' \emph{Journal of Autism and
  Developmental Disorders}, vol.~28, no.~3, pp. 189--198, 1998.

\bibitem{koegel2009improving}
R.~L. Koegel, T.~W. Vernon, and L.~K. Koegel, ``Improving social initiations in
  young children with autism using reinforcers with embedded social
  interactions,'' \emph{Journal of autism and developmental disorders},
  vol.~39, no.~9, pp. 1240--1251, 2009.

\bibitem{lord2000autism}
C.~Lord, S.~Risi, L.~Lambrecht, E.~H. Cook, B.~L. Leventhal, P.~C. DiLavore,
  A.~Pickles, and M.~Rutter, ``The autism diagnostic observation
  schedule—generic: A standard measure of social and communication deficits
  associated with the spectrum of autism,'' \emph{Journal of autism and
  developmental disorders}, vol.~30, no.~3, pp. 205--223, 2000.

\bibitem{lord2022lancet}
C.~Lord, T.~Charman, A.~Havdahl, P.~Carbone, E.~Anagnostou, B.~Boyd, T.~Carr,
  P.~J. De~Vries, C.~Dissanayake, G.~Divan \emph{et~al.}, ``The lancet
  commission on the future of care and clinical research in autism,'' \emph{The
  Lancet}, vol. 399, no. 10321, pp. 271--334, 2022.

\bibitem{oakland2011adaptive}
T.~Oakland and P.~L. Harrison, \emph{Adaptive behavior assessment system-II:
  Clinical use and interpretation}.\hskip 1em plus 0.5em minus 0.4em\relax
  Academic Press, 2011.

\bibitem{oakland2008adaptive}
------, ``Adaptive behaviors and skills: An introduction,'' in \emph{Adaptive
  Behavior Assessment System-II}.\hskip 1em plus 0.5em minus 0.4em\relax
  Academic Press, 2008, pp. 1--20.

\bibitem{wallace2012global}
S.~Wallace, D.~Fein, M.~Rosanoff, G.~Dawson, S.~Hossain, L.~Brennan, A.~Como,
  and A.~Shih, ``A global public health strategy for autism spectrum
  disorders,'' \emph{Autism Research}, vol.~5, no.~3, pp. 211--217, 2012.

\bibitem{kasari2013interventions}
C.~Kasari and T.~Smith, ``Interventions in schools for children with autism
  spectrum disorder: Methods and recommendations,'' \emph{Autism}, vol.~17,
  no.~3, pp. 254--267, 2013.

\bibitem{mullen2011mastering}
T.~Mullen, \emph{Mastering blender}.\hskip 1em plus 0.5em minus 0.4em\relax
  John Wiley \& Sons, 2011.

\bibitem{deliyski1993acoustic}
D.~D. Deliyski, ``Acoustic model and evaluation of pathological voice
  production,'' in \emph{Third European Conference on Speech Communication and
  Technology}, 1993.

\bibitem{kang2013effects}
S.~Kang, M.~O’Reilly, L.~Rojeski, K.~Blenden, Z.~Xu, T.~Davis, J.~Sigafoos,
  and G.~Lancioni, ``Effects of tangible and social reinforcers on skill
  acquisition, stereotyped behavior, and task engagement in three children with
  autism spectrum disorders,'' \emph{Research in developmental disabilities},
  vol.~34, no.~2, pp. 739--744, 2013.

\end{thebibliography}

\end{document}